\documentclass[twocolumn,showpacs,preprintnumbers,superscriptaddress,prm,floatfix,aps,10pt]{revtex4-2}
\usepackage[utf8]{inputenc}
\usepackage{natbib}
\usepackage{amsmath,amssymb}
\usepackage{xcolor}
\usepackage{todonotes}
\usepackage{comment}
\usepackage{xr-hyper}
\usepackage{url}
\usepackage{hyperref}
\hypersetup{
    colorlinks=true,
    citecolor=blue,
    linkcolor=blue,
    filecolor=magenta,      
    urlcolor=blue,
    pdfpagemode=FullScreen,
    }
\definecolor{olle}{RGB}{240,226,182}
\definecolor{nimrod}{RGB}{192,225,215}
\definecolor{omer}{RGB}{192,150,215}

\newcommand{\wn}{cm$\textsuperscript{-1}$}
\renewcommand {\vec}    [1]    {\ensuremath{\boldsymbol{#1}}}

\makeatletter

\newcommand*{\rom}[1]{\expandafter\@slowromancap\romannumeral #1@}

\begin{document}

\title{The dielectric response of rock-salt crystals at finite temperatures from first principles}
\date{\today}

\author{Nimrod Benshalom}
\affiliation{Department of Chemical and Biological Physics, Weizmann Institute of Science, Rehovot 76100, Israel}
\author{Guy Reuveni}
\affiliation{Department of Chemical and Biological Physics, Weizmann Institute of Science, Rehovot 76100, Israel}
\author{Roman Korobko}
\affiliation{Department of Chemical and Biological Physics, Weizmann Institute of Science, Rehovot 76100, Israel}
\author{Omer Yaffe}
\email{omer.yaffe@weizmann.ac.il}
\affiliation{Department of Chemical and Biological Physics, Weizmann Institute of Science, Rehovot 76100, Israel}
\author{Olle Hellman}
\email{olle.hellman@liu.se}
\affiliation{Department of Physics, Chemistry and Biology (IFM), Link\"oping University, SE-581 83, Link\"oping, Sweden.}
\affiliation{Department of Molecular Chemistry and Material Science, Weizmann Institute of Science, Rehovot 76100, Israel}

\begin{abstract}
We combine \emph{ab initio} simulations and Raman scattering measurements to demonstrate explicit anharmonic effects in the temperature dependent dielectric response of a NaCl single crystal.
We measure the temperature evolution of its Raman spectrum and compare it to both a quasi-harmonic and anharmonic model.
Results demonstrate the necessity of including anharmonic lattice dynamics to explain the dielectric response of NaCl, as it is manifested in Raman scattering.
Our model fully captures the linear dielectric response of a crystal at finite temperatures and may therefore be used to calculate the temperature dependence of other material properties governed by it.
\end{abstract}

\maketitle

\section{Introduction}

The microscopic manifestation of temperature in lattice dynamics is generally understood through the \textit{harmonic approximation}, where atomic motion is mapped onto a set of vibrational normal modes (i.e., phonons)~\cite{Born1998}. 
Modern studies into lattice dynamics of solids significantly expand on this view to account for anharmonic phenomena such as phase transitions~\cite{Fultz2020,Bruce1981}, thermal expansion~\cite{Barron1999} and thermal conductance~\cite{Sun2010,Romero2015,Whalley2016}.

In contrast, studies that focus on the electronic properties of solids rarely go beyond a quasi-harmonic (QH) treatment, where the normal modes are re-normalized for each given temperature to account for thermal expansion~\cite{Fultz2010}.
The effects of lattice dynamics are usually incorporated through the prism of electron-phonon interaction. 
This can manifest in modified carrier lifetimes~\cite{Eiguren2002,Eiguren2003}, corrections to the electronic band structure~\cite{Gopalan1987}, mobility estimations~\cite{Li2015,Fiorentini2016,Schweicher2019} and calculations for vibrational spectroscopy~\cite{Strauch1995,Gillet2017,Cusco2007,Menedez1983}.
Formally, almost any \emph{ab initio} calculation of electron-phonon interactions relies on the harmonic approximation~\cite{Giustino2017}.
A different manifestation of lattice dynamics in electronic properties is the temperature evolution of the material's dielectric response. 
The need to extend anharmonic modeling to dielectric properties has been demonstrated perhaps most strikingly in halide perovskite semiconductors~\cite{Zhu2016,Joshi2019,Martiradonna2018,Sender2016,Schilcher2021,Guo2019,Omer2017}.
Importantly, by relying on molecular dynamics simulations, calculations of the dielectric response of a crystal are not limited to a perturbative treatment of anharmonicity.

Because inelastic light scattering originates in the vibrational modulation of a crystal's polarizability auto-correlation function~\cite{Yu2010}, it is the ideal probe to study the effect of anharmonicity on dielectric properties.
A main drawback of optical inelastic light scattering is that momentum conservation limits vibrational contributions to the $\Gamma$, or zero crystal momentum point. 
This limitation does not apply in 2\textsuperscript{nd} order Raman scattering, which is driven by contributions from the entire Brillouin zone (BZ)~\cite{Loudon1964}. 
Simply put, 2\textsuperscript{nd} order Raman describes the inelastic scattering of a single photon with two phonons~\cite{Cardona1982}. 
The scattering intensity inside a given frequency interval $I\left(\Omega\right)\text{d}\Omega$ emerges out of contributions from all possible phonon combinations with the appropriate energy and momentum conservation. 
It is therefore sensitive to any anharmonic effect in either the phonon dispersion or the dielectric response.

In this study, we introduce an \textit{ab initio} approach incorporating finite temperature effects into the dielectric response of a crystal. 
The approach is based on the effective interatomic force constants used in the temperature dependent effective potential (TDEP) method~\cite{Hellman2011,Hellman2013a,Hellman2013}. 
The computational method is benchmarked with experimental measurements of dielectric response in terms of 2\textsuperscript{nd} order Raman scattering.
As a model system we choose NaCl.
It is an ideal showcase since first order Raman is forbidden by symmetry, necessitating a computational methodology that goes beyond the lowest order.
Furthermore, the central role of anharmonicity in NaBr was recently demonstrated~\cite{Shen2020}, motivating further examination of the rock-salt structure.
The comparison between simulated and experimental results shows that the harmonic treatment is inadequate when considering dielectric response at finite temperatures.
In contrast, the method developed here not only reproduces experimental results more faithfully than conventional approaches, it also shows how higher order dielectric response is linked to other manifestations of anharmonicity.
More specifically, it accounts for phonon broadening, thermal transport, and demonstrates how a measurement of the 2\textsuperscript{nd} order Raman spectrum can be used to conclude the dominant anharmonic expressions in a given material.
Importantly, due to its \emph{ab initio} nature, our method provides a full description of the linear dielectric response of the material and its dependence on atomic displacements.

\begin{figure}
\centering
\includegraphics[width=\linewidth]{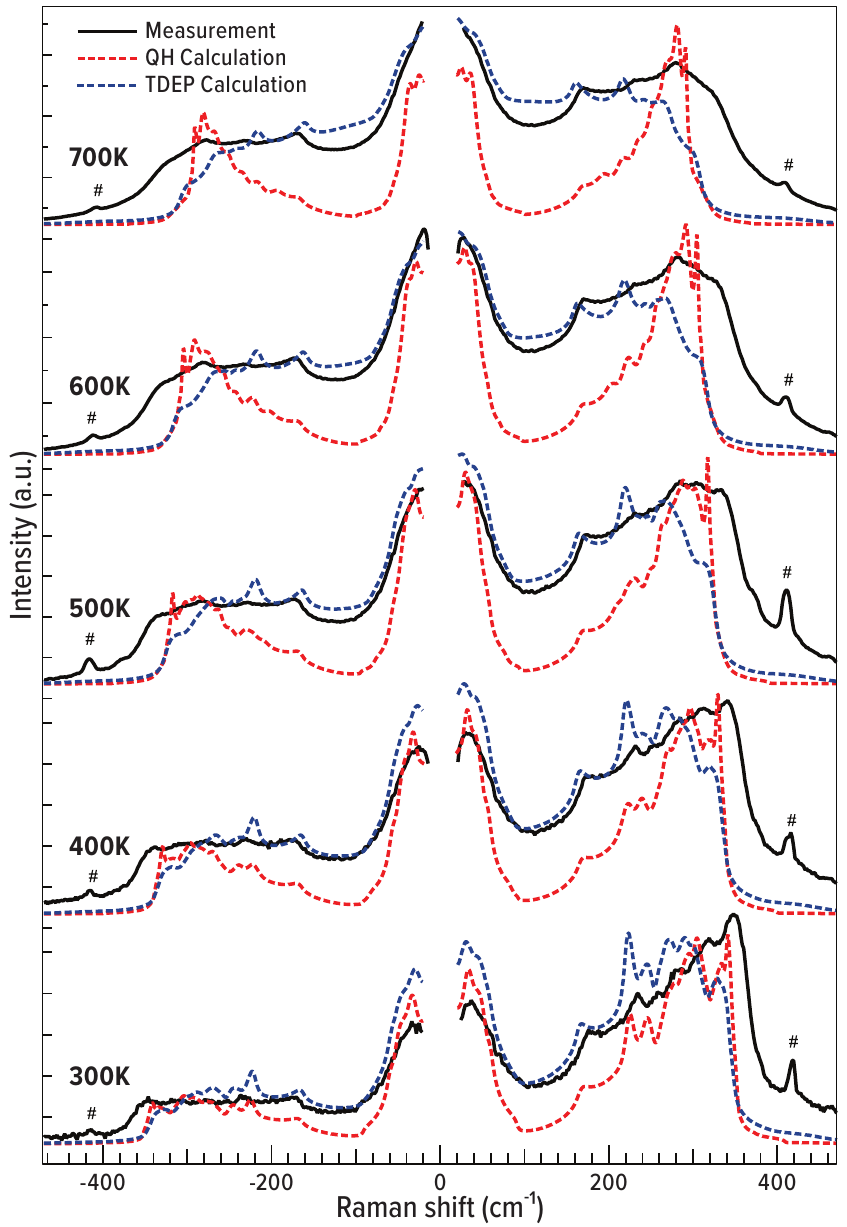}
\caption{Temperature dependence of the Raman spectrum for NaCl by measurement (\textbf{black}), by quasi-harmonic (\textcolor{red}{red}) and TDEP (\textcolor{blue}{blue}) simulation. \# symbols mark system artifacts.}
\label{fig:NaClTemp}
\end{figure}

\section{The Raman spectrum of Sodium Chloride}

\subsection{Experimental and calculated spectra}

Our main results are presented in Fig.~\ref{fig:NaClTemp}, where we compare the measured, temperature dependent, unpolarized \footnote{measured/calculated sums of the scattered intensities for a full rotation of incident polarization} Raman spectra of NaCl and corresponding simulations. 
The technical details for the construction of an unpolarized spectrum, as well as the QH computation, are provided in Sec.~S\rom{2},~S\rom{7}, respectively, in the supplementary material (SM). 
Our experimental spectra are in good agreement with past measurements~\cite{Welsh1949, Krauzman1969}. 
We observe a continuous spectrum with some sharp non-Lorentzian features centered around 280 cm$^{-1}$.

Unlike early analytical treatments~\cite{Birman1962,Birman1963,Lax1961}, our computation is able to predict the entire shape of the spectrum. 
It is immediately evident that even at 300~K the QH calculation for the spectrum is imperfect.
It overestimates the high ($\approx$300~\wn) to low ($\approx$100~\wn) intensity ratio.
It also fails to capture the broad decaying intensity above 350~\wn.
The discrepancies between experiment and QH calculation at 300~K are remarkable, as they demonstrate the significance of anharmonic effects in the dielectric response of an archetypal rock-salt crystal, even at room temperature. 
These discrepancies get even larger at higher temperatures.
Above 500~K the main spectral feature around 300~cm$\textsuperscript{-1}$ broadens and flattens but the QH calculation predicts an opposite trend. 
These findings constitute direct evidence for the failure of a QH approach to explain the dielectric response of NaCl, due to finite temperature anharmonic effects.

Contrary to the QH approach, the TDEP-based computation shows much better agreement with experiment. 
At 300~K the line-shape of the TDEP-based Raman spectrum (blue) follows more accurately the experimental spectrum (black). 
Agreement between experimental and TDEP spectra is even better above 500~K. 
The main discrepancy remains the decaying intensity above 350~cm$\textsuperscript{-1}$ which is not captured by either QH or TDEP. 
We suspect this is the result of higher order ($>2$) scattering terms since, as discussed below, our calculation does not include them. 
Having established that our computational method is both required and successful in predicting the Raman spectrum, we turn to discuss the TDEP-based calculation in more detail.


\subsection{Dielectric response of a crystal from first principles}

We start by briefly reiterating the main idea behind TDEP to describe the ionic motion. 
The starting point is the lattice dynamical expansion for the crystal energy:~\cite{Born1998}
\begin{equation}
\begin{split}
    H = & H_0 + 
    \frac{1}{2}
    \sum_{ij\alpha\beta} \Phi_{ij}^{\alpha\beta} 
    u_i^{\alpha} u_j^{\beta}
     +
    \frac{1}{3!}
    \sum_{ikj\alpha\beta\gamma} \Phi_{ijk}^{\alpha\beta\gamma} 
    u_i^{\alpha} u_j^{\beta} u_k^{\gamma} +
\\ &    
    +
    \frac{1}{4!}
    \sum_{ikjl\alpha\beta\gamma\delta} \Phi_{ijkl}^{\alpha\beta\gamma\delta} 
    u_i^{\alpha} u_j^{\beta} u_k^{\gamma} u_l^{\delta}
    +
    \ldots
\end{split}
\end{equation}
where $u_i^\alpha$ denote the displacement of atom $i$ in direction $\alpha$. The interatomic force constants $\Phi$ are determined by minimizing the difference between the model system and \emph{ab initio} calculated forces $\vec{f}$.
\begin{equation}
    \vec{\Phi} = \arg\min_{\Phi} \| \vec{f}^{\textrm{model}} - \vec{f}^{\textrm{ab initio}} \|
\end{equation}
The \emph{ab initio} calculations are performed in the canonical ensemble yielding an explicit temperature dependent model Hamiltonian -- either via molecular dynamics or a self-consistent stochastic sampling~\cite{Shulumba2017}. 
In this study we used the latter.

Since the sampling is done at finite temperature, an explicit temperature dependence is built into the effective interaction parameters. The TDEP Hamiltonian is constructed to be the best possible fit at a given temperature.

To asses the accuracy of our model Hamiltonian for the lattice dynamics we calculate the phonon spectral function and compare it with a neutron scattering experiment~\cite{Raunio1969a}. 
The calculated spectral functions agree well with experiments and previous non-harmonic calculations~\cite{Ravichandran2018} (see Sec.~S\rom{8} in the SM).

The need for a non-perturbative treatment when describing anharmonic lattice dynamics motivates an analogous treatment for the dielectric response. 
To extend the formalism to dielectric properties we similarly expand the susceptibility $\vec{P}$ and dipole moment $\vec{M}$ in terms of atomic displacements $u$:~\cite{Born1998}
\begin{align}
\begin{split}
    \label{eq:pexpansion}
    P^{\mu\nu} = & P_0^{\mu\nu} + 
    \sum_{i\alpha}
    P^{\mu\nu\alpha}_i u_i^{\alpha} +
    \frac{1}{2}
    \sum_{ij\alpha\beta}
    P^{\mu\nu\alpha\beta}_{ij} u_i^{\alpha}u_j^{\beta} + 
    \ldots
\end{split}
\\
\begin{split}
    \label{eq:mexpansion}
    M^{\mu} = & M_0^{\mu} + 
    \sum_{i\alpha}
    M^{\mu\alpha}_i u_i^{\alpha} + 
    \frac{1}{2}
    \sum_{ij\alpha\beta}
    M^{\mu\alpha\beta}_{ij} u_i^{\alpha}u_j^{\beta} + 
    \ldots
\end{split}
\end{align}
Here the indices $\mu\nu$ denote Cartesian components of electric field derivatives. 
In principle this expansion is done for dynamic quantities -- including the electronic frequency dependence -- but in the present case for NaCl we are interested in a frequency range far from any resonances and can safely ignore the electronic frequency dependence.

In analogy with the process described for the force constants, the terms in the dipole moment and susceptibility expansions are determined by minimizing the difference between the model values and those obtained from simulation:
\begin{align}
    \vec{M} & = \arg\min_{\vec{M}} \left\| 
        \frac{\partial \vec{M}}{\partial u}^{\textrm{model}} - 
        \frac{\partial \vec{M}}{\partial u}^{\textrm{ab initio}}
    \right\|
\\
    \vec{P} & = \arg\min_{\vec{P}} \left\| 
        \vec{P}^{\textrm{model}} - 
        \vec{P}^{\textrm{ab initio}}
    \right\|.
\end{align}

The end result is a set of interaction parameters that explicitly depend on temperature, incorporating all orders of non-harmonic effects in the linear dielectric response of a material. 
We note that the inclusion of frequency dependence adds no conceptual difficulty, only the practical challenge of accurately determining the frequency dependent electronic dielectric response for a simulation with hundreds of atoms. 
The algorithm for determining the interaction parameters is described in detail in Sec.~S\rom{5} in the SM, and the specific computational details are given in Sec.~S\rom{8}.
Having obtained the dielectric response we turn to compute the Raman scattering cross-section.
\begin{figure*}[t]
\centering
\includegraphics[width=\linewidth]{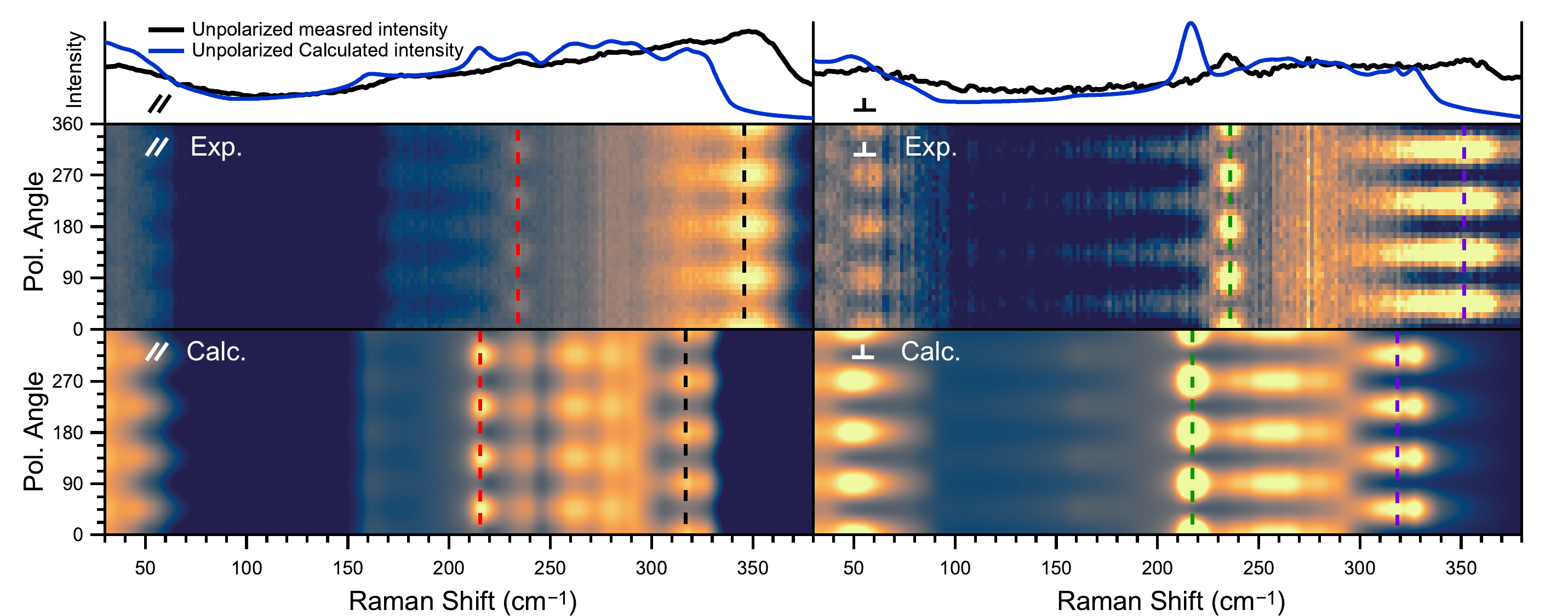}
\caption{NaCl Raman PO color map for measurement and TDEP simulation in 300~K. Frequency of spectral features in simulation slightly misses experimental values, so matching dashed color lines help guide comparison.
}
\label{fig:NaClPO}
\end{figure*}
\subsection{\textit{Ab initio} calculation of the Raman spectrum of a crystal} 


Theoretical interpretations of 2\textsuperscript{nd} order Raman spectra usually limit themselves to high symmetry areas of reciprocal space where most density of states singularities are expected~\cite{Burstein1965,Agrawal1967,Karo1965,Weber1993,Wang1973,Smith2002}.
This approach adequately accounts for pronounced peaks, but severely limits the possibility of incorporating anharmonic effects.
Also, because of the symmetry breaking of moving away from the $\Gamma$-point and generalization into multiple phonon states, selection rules in 2\textsuperscript{nd} order Raman usually become too relaxed to place meaningful restrictions on the measured Raman tensor (see Sec.~S\rom{4} in SM for more details).
Existing \textit{ab initio} approaches to 2\textsuperscript{nd} order Raman all operated strictly within the harmonic approximation~\cite{Gillet2017,Strauch1995}.
A Shell-model based calculation by \citet{Bruce1972} did incorporate phonon-phonon interactions, but his partial treatment lead him to conclude these are negligible in the Raman spectrum of NaCl.

The construction of the full 2\textsuperscript{nd} order spectrum from first principles requires a few computational steps.
The starting point is to use the TDEP method to construct an anharmonic Hamiltonian that describes the dynamics of the ions.
This anharmonic Hamiltonian is then coupled to an expansion of the polarizability in terms of atomic displacements.
The coupled ionic-polarizability system can then be solved with many-body methods, providing the finite temperature spectrum.
The steps involved are discussed in detail below.

Following \citet{Cowley1963}, the Raman scattering cross-section is given by:
\begin{equation}
    \sigma(\Omega) \propto 
    \sum_{\mu\nu\xi\rho}
	E^{\textrm{out}}_{\mu}
	E^{\textrm{out}}_{\xi}
	I_{\mu\nu,\xi\rho}(\Omega)
	E^{\textrm{in}}_{\nu}
	E^{\textrm{in}}_{\rho}    
\end{equation}
where $\vec{E}^{\textrm{in}}$ and $\vec{E}^{\textrm{out}}$ are the electric field vectors of the incoming and outgoing light, with Greek letters standing for Cartesian components.
The tensor $I$ depends on the probing frequency $\Omega$ and constitutes the material property governing Raman scattering. 
It is given by
\begin{equation}
\label{eq:Cowley_Beast}
    I_{\mu\nu,\xi\rho}(\Omega) = \int \text{d}t \left\langle P^{\mu\nu}(t)P^{\xi\rho}(0) \right\rangle e^{-i\Omega t},
\end{equation}
the Fourier transformed thermal average of the polarizability-polarizability autocorrelation function.
Eq.~\eqref{eq:Cowley_Beast} says that the complete description for Raman scattering is given by a tensor detailing the vectorial relationship between incident and scattered fields.
It is therefore desirable to inspect polarization dependence when comparing theory and measurement.

To realize this, we performed a polarization-orientation (PO) Raman measurement of a NaCl single crystal. 
The temperature dependent spectra presented in Fig.~\ref{fig:NaClTemp} are actually calculated sums of the scattered intensities for a full rotation of incident polarization.
We probe separately the parallel and perpendicular scattered polarizations with respect to the linearly polarized incident beam orientation.
Details of the experimental setup are given in Sec.~S\rom{1} in the SM.

Figure~\ref{fig:NaClPO} shows the full PO color map for measurement and simulation in 400~K (PO maps for all temperatures are given in Sec.~S\rom{2} in the SM).
Our theoretical calculation also fully accounts for the tensor nature of Raman scattering, making a detailed comparison possible.
The calculated model closely follows the PO dependence throughout the spectrum.
Dashed colored lines follow corresponding spectral features in simulation and experiment (some discrepancy in absolute frequency persists).
Our calculation correctly predicts the periodicity with incident angle of each feature, as well as the relative phase between them, thereby faithfully reproducing the tensorial nature of the modulated dielectric response.

The full tensorial formalism for 2\textsuperscript{nd} order Raman has already been rigorously worked out, but is rarely used and is repeated and extended here~\cite{Cowley1964b,Wallis1971}.
Far from resonance, to lowest order, the tensor $I$ relating incident and outgoing light has four terms:
\begin{equation}
    I_{\mu\nu,\xi\rho}(\Omega) = 
    I^{(\textrm{I})}_{\mu\nu,\xi\rho}+
    I^{(\textrm{II})}_{\mu\nu,\xi\rho}+
    I^{(\textrm{III})}_{\mu\nu,\xi\rho}+
    I^{(\textrm{IV})}_{\mu\nu,\xi\rho}
\end{equation}
with the different terms given by
\begin{align}
    \label{eq:ramanfirstorder}
	I^{(\textrm{I})}_{\mu\nu,\xi\rho} = &
	(n(\Omega)+1)
	\sum_s P^{(\textrm{I})}_{\mu\nu,\xi\rho}(s) 
	J_s(\Omega)    
\\
\label{eq:ramansecondorder}
\begin{split}
	I^{(\textrm{II})}_{\mu\nu,\xi\rho} = &
	2
	\sum_{ \vec{q} s_1 s_2 }
	P^{(\textrm{II})}_{\mu\nu,\xi\rho}(\vec{q},s_1,s_2)
		\int 
	J_{s_1}(\Omega')(n(\Omega')+1)
 \times \\
	&
	J_{s_2}(\Omega-\Omega')(n(\Omega-\Omega')+1)
	d\Omega'
\end{split}
\\
\label{eq:ramanIII}
\begin{split}
	I^{(\textrm{III})}_{\mu\nu,\xi\rho} = & 
	-6(n(\Omega)+1)
	\sum_{\vec{q}s_1 s_2 s_3}
	J_{s_3}(\Omega)
	\times \\
	& P^{(\textrm{III})}_{\mu\nu,\xi\rho}(\vec{q},s_1,s_2,s_3)
	\,\Im\left\{ S(s_1,s_2,\Omega) \right\}
	\vspace{5pt}
\end{split}
\\
\label{eq:ramanIV}
\begin{split}
	I^{(\textrm{IV})}_{\mu\nu,\xi\rho} = & 
	3(n(\Omega)+1)
	\sum_{\vec{q} s_1 s_2}
	P^{(\textrm{IV})}_{\mu\nu,\xi\rho}(\vec{q},s_1,s_2) \times
	\\ & 
	(2n_{s_2}+1)
	J_{s_1}(\Omega)    
\end{split}
\end{align}
where the matrix elements $\vec{P}^{(\textrm{I})}-\vec{P}^{(\textrm{IV})}$ are defined via the Fourier components of the terms in Eq.~\eqref{eq:pexpansion}, (see Sec.~S\rom{7} in the SM for the explicit expressions and derivations of the above terms). 
The phonon spectral function for mode $s$, $J_s = -\Im\left\{ G_{s}(\Omega) \right\}/\pi$ is obtained from
\begin{equation}
    G_{s}(Z) = \frac{2\omega_s}{\omega_s^2 - 2\omega_s\Sigma_s(Z) - Z^2}
\end{equation}    
where
\begin{equation}
\begin{split}
    \Sigma_{\vec{q}s}(Z) = & -18 \sum_{\vec{q}_1\vec{q}_2 s_1 s_2}
    \left| \Phi^{s s_1s_2}_{\vec{q}\vec{q}_1\vec{q}_2} \right|^2
    S(s_1,s_2,Z) +
\\ & +
    12 \sum_{\vec{q}_1 s_1}
    \Phi^{s s s_1 s_1}_{\vec{q}\bar{\vec{q}}\vec{q}_1\bar{\vec{q}}_1}(2n_{\vec{q}_1 s_1}+1)
\end{split}    
\end{equation}
and
\begin{equation}
\label{eq:Sfun}
\begin{split}
    &S(s_a,s_b,Z) = 
    \\
	&(n_{a}+n_{b}+1)
	\left[
	\frac{1}{(\omega_{a}+\omega_{b}-Z)_p}-
	\frac{1}{(\omega_{a}+\omega_{b}+Z)_p}
	\right]
	\\
	&+
	(n_{a}-n_{b})
	\left[
	\frac{1}{(\omega_{b}-\omega_{a}+Z)_p}-
	\frac{1}{(\omega_{b}-\omega_{a}-Z)_p}
	\right]
\end{split}	
\end{equation}
The first term, Eq.~\eqref{eq:ramanfirstorder}, is the first order Raman scattering which comes down to the one-phonon spectral function weighted by the first order Raman matrix elements. 
The spectral function contains any broadening and shifts due to anharmonicity, and also any deviation from a Lorentzian lineshape. 
This term is what is commonly referred to as Raman scattering.

The second term, Eq.~\eqref{eq:ramansecondorder}, is the second order Raman scattering. 
The matrix elements contain the momentum conservation, and the convolution term contains the energy conservation. 
If one were to set the matrix elements to unity it would yield a spectrum that follows the two-phonon density of states.

The third and fourth terms are more subtle, as they contain contributions from the whole BZ but are multiplied with the one-phonon lineshape. 
They will thus decay rapidly away from the first order Raman peaks, with the net effect of slightly shifting and altering the shapes of the first order peaks. 
These terms are one reason that neutron scattering and Raman scattering measurements might not coincide exactly. 

The theoretical PO maps in Fig.~\ref{fig:NaClPO} are calculated from Eq.~\eqref{eq:ramansecondorder}. In NaCl first order Raman is forbidden by symmetry, which means $P^{(\textrm{I})}$, $P^{(\textrm{III})}$, and $P^{(\textrm{IV})}$ are zero. We stress that the spectrum is calculated from the interacting phonons, i.e., a convolution of one-phonon spectral functions and includes explicit anharmonic effects.

\begin{figure}[h]
    \centering
    \includegraphics{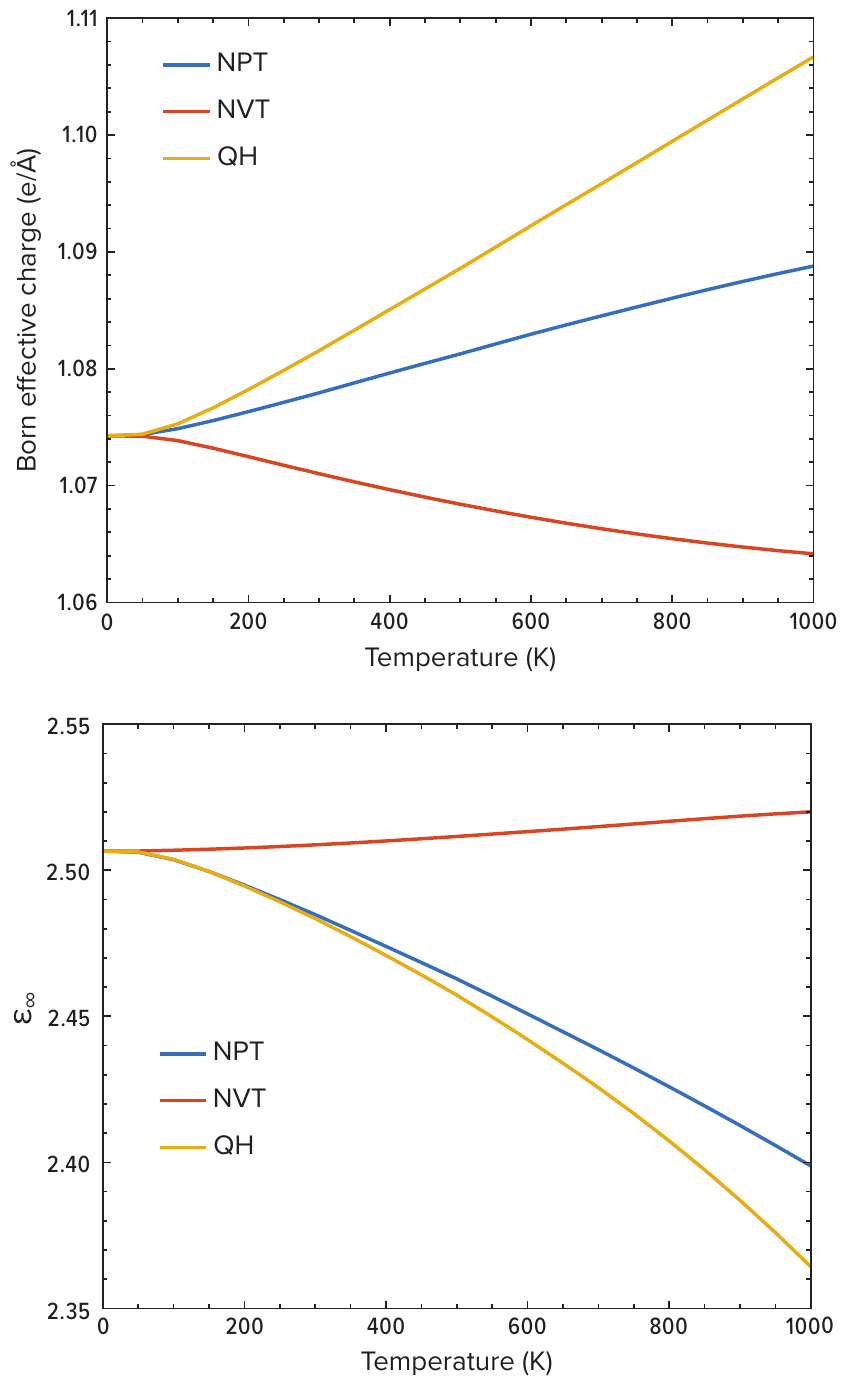}
    \caption{Temperature dependence of the Born effective charge and $\epsilon_\infty$ in NaCl treated in three different ways. 
    The most realistic, at constant pressure (i.e. volume changing with temperature) with full temperature dependence of all parameters is denoted NPT. The same treatment without thermal expansion is denoted NVT. 
    For reference we also include the quasiharmonic results, denoted QH, where all interaction parameters are set to their values at 0K, and temperature is only included via thermal expansion.}
    \label{fig:SMZandeps}
\end{figure}

\begin{figure*}[ht]
    \centering
    \includegraphics[width=\linewidth]{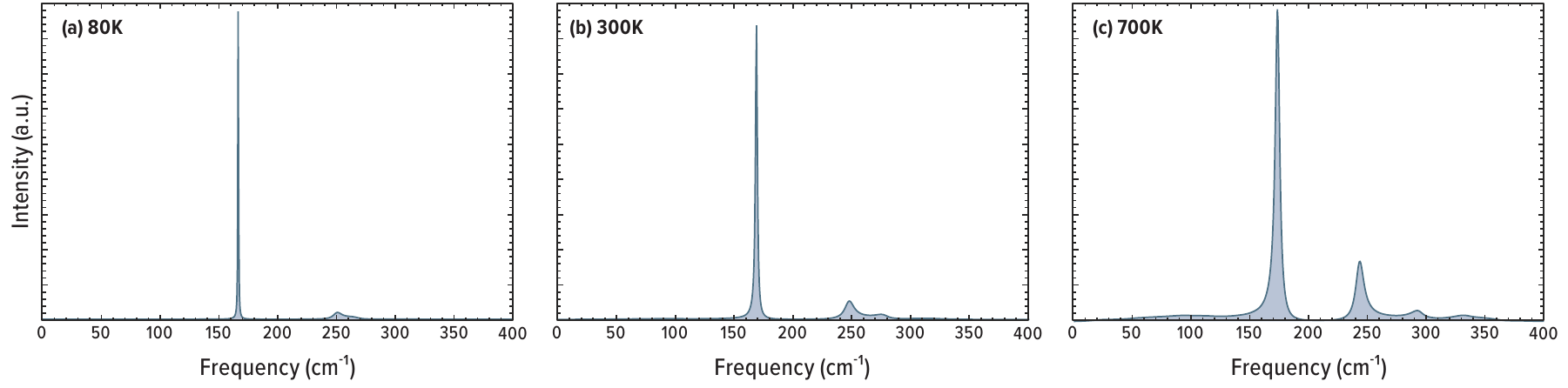}
    \caption{Temperature dependence of the IR absorption spectra in NaCl. The spectra are calculated via equations \eqref{eq:ramanfirstorder}--\eqref{eq:ramanIV} with the Raman matrix elements replaced with the IR matrix elements, equations S83--S86 in Sec.~S\rom{2} in the SM.}  
    \label{fig:nacl_Z_and_eps}
\end{figure*}

\subsection{Anharmonic effects in 2nd order Raman}

It is worth dwelling on the temperature effects in 2\textsuperscript{nd} order Raman, and their connection to anharmonic behavior.
The scattering is generated by phonons throughout the BZ. 
Since the TDEP method implicitly contains all orders of anharmonicity, this already imparts a non-trivial temperature dependence, but the effects of anharmonicity go beyond the temperature dependence of the bare phonon dispersion curve. 
Inspecting Eq.~\eqref{eq:ramansecondorder} one sees that it is a convolution of two one-phonon spectral functions. 
If we replace the interacting spectral functions with the non-interacting ones we will recover the traditional result~\cite{Cowley1964b} (See SM, Eq.~S73).
In the present study we found it crucial to include the interacting spectral functions when calculating the 2\textsuperscript{nd} order Raman spectra, without them agreement with experiment was markedly worse. 
Intuitively, if the one-phonon DOS is broadened, the two-phonon DOS must also broaden -- and since the two-phonon DOS is a convolution, non-Lorentzian lineshapes have a larger impact on two-phonon than on one-phonon spectra.

Besides its own non-trivial temperature dependence, 2\textsuperscript{nd} order Raman emerged as a uniquely sensitive probe for anharmonic lattice dynamics. 
Three-phonon scattering is ostensibly a straightforward perturbative anharmonic effect.
Only a small fraction of all possible three-phonon combinations satisfy energy and momentum conversion -- this is what we refer to as the scattering phase space. 
Mathematically, this is defined via the imaginary part of Eq.~\eqref{eq:Sfun}. 
The temperature dependence of available phonon scattering phase space is, however, a non-perturbative anharmonic effect, and can lead to a host of surprising trends with temperature.
Thermal conductivity that does not follow $T^{-1}$~\cite{Romero2015}, phonon line widths that do not grow linearly with temperature~\cite{Delaire2011b}, or in our case, a Raman spectrum that does not follow the quasiharmonic approximation, are all consequences of non-pertrubative anharmonicity.

The 2\textsuperscript{nd} order Raman spectrum follows the same set of momentum and energy conservation rules as three-phonon scattering.
In fact, the spectra is directly proportional to the available scattering phase space. 
This makes Raman scattering invaluable as a direct probe of anharmonicity: we can see how (a part) of the scattering phase space evolves with temperature, providing insight into the underlying mechanisms governing strongly anharmonic materials.
In this specific case we can verify prior theoretical studies. \citet{Ravichandran2018} found it necessary to include temperature-dependent phonons to accurately describe the thermal transport in NaCl -- this can be verified experimentally by the change in the Raman spectra with temperature in Fig.~\ref{fig:NaClTemp}. Moreover, four-phonon scattering was predicted to be important. The presence of higher order scattering is evident in the tail at large wavenumbers (second order Raman can only contribute up to wavenumbers equal to $2\omega_{\textrm{max}}$, third order Raman, which would involve four-phonon scattering, can contribute up to $3\omega_{\textrm{max}}$).

\section{Temperature dependence of additional dielectric properties}

The methodology described in this letter is general with regard to the interplay between anharmonicity and dielectric response. 
Experimentally, we have the Raman spectra to verify our results with. 
To illustrate the general nature of the method we calculate a few other quantities accessible by it.
We present the infrared absorption spectra, as well as the temperature dependence of the dielectric tensor and Born charges. 
We could not find any experimental results to verify these quantities with, so we leave them as predictions.

In Fig.~\ref{fig:SMZandeps} we show the temperature dependence of the Born charges and static dielectric tensor. 
The temperature dependence is modest as one would expect in a wide-bandgap material. 
Even so, the quasiharmonic treatment of Born charges overestimates the temperature dependence by a factor of 2 when compared with the TDEP calculation. 
It is easy to imagine that in a material with a much smaller bandgap this discrepancy could be significantly larger.

In Fig.~\ref{fig:nacl_Z_and_eps} we show the IR spectra of NaCl for a few temperatures.
The procedure to calculate these is identical to the Raman spectra, the only difference being the matrix elements.


\section{Conclusions}

In conclusion, we have demonstrated the crucial role of anharmonic lattice dynamics effects in determining the phonon-modulated linear dielectric response of a crystal. 
This was achieved by introducing an \textit{ab initio} method to calculate the dielectric response of a crystal at finite temperatures and comparing it to the first continuous PO spectrum of an exclusively 2\textsuperscript{nd} order Raman structure.
Our generalized TDEP method incorporates all orders of non-harmonic effects through a sampling of temperature dependent lattice configurations, expressed as effective interatomic force constants.
We introduce analogous effective interaction tensors that govern the dielectric response, i.e., we expand the dipole moment and polarizability in terms of atomic displacements.
Using the new generalized TDEP method, other dielectric material properties such as ferroelectricity and capacitance may be calculated, along with their temperature dependence. 
By isolating and evaluating the normal modes comprising the emergent temperature dependence of a material property, new design rules may be inferred for better functional materials operating at finite temperatures.

\section*{acknowledgments}
The authors would like to thank Dr. Lior Segev (WIS) for invaluable software development. O.Y. acknowledges funding from ISF(1861/17), BSF (grant no. 2016650) and ERC (850041 -ANHARMONIC). O.H. acknowledges support from the Swedish Research Council (VR) program 2020-04630. Supercomputer resources were provided by the Swedish National Infrastructure for Computing (SNIC).  

\bibliography{main.bib}

\end{document}


\title{Project hiding frog, the hidinger stuff.}
\begin{center}
    \large{\textbf{The dielectric response of rock-salt crystals at finite temperatures from first principles}} \\
    \LARGE{Supplementary Material}
\end{center}


\section{Experimental setup}  \label{SM:Experiment}
\begin{figure}
    \centering
    \includegraphics[width=\textwidth]{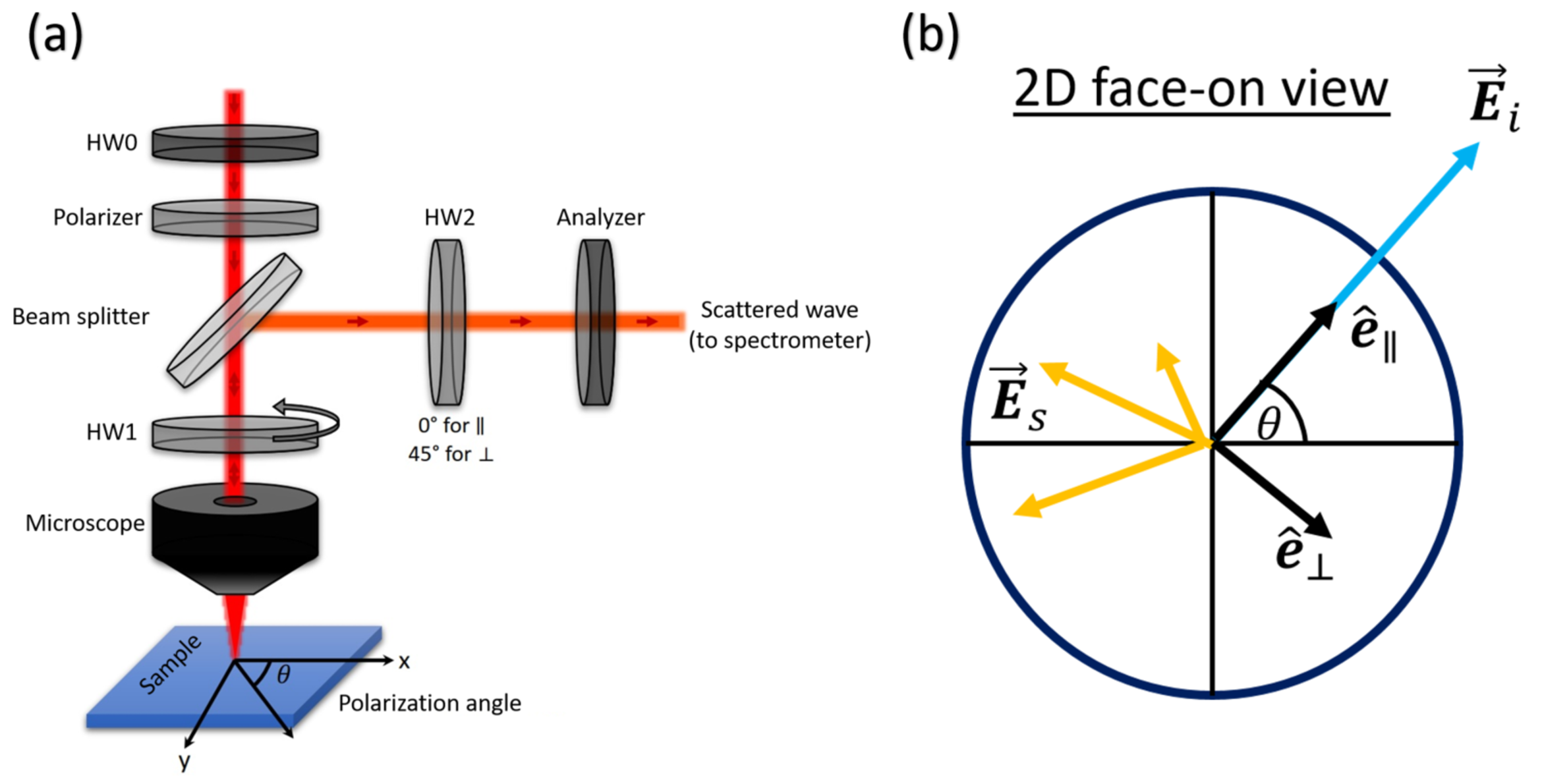}
    \caption{({\bfseries a}) Schematic of experimental setup. Laser light comes from top. Scattered signal goes right for detection. 
    ({\bfseries b}) Diagram for the different polarizations involved in the measurement. $\overset{\rightarrow}{\vec{E}_i}$ is the incident linear polarization. $\overset{\rightarrow}{\vec{E}_s}$ is the polarization for the scattered signal and need not be linearly polarized.
    Parallel measurement (HW2 at 0$\textsuperscript{0}$) 
    probes the scattered field's projection on $\hat{\vec{e}}_{\parallel}$. 
    The perpendicular measurement (HW2 at 45\textsuperscript{0}) 
    probes the scattered field's projection on $ \hat{\vec{e}}_{\perp} $.
    \label{fig:SM_setup}}
\end{figure}

All Raman measurements were performed on our home-built system. Figure \ref{fig:SM_setup} shows a schematic of the system. 
The beam path begins with a 488nm solid-state (Coherent Sapphire SF 488-100 CDRH) laser, which will excite the Raman scattering in the sample. 
The laser is filtered for any amplified spontaneous emission by a volume holographic ASE (Ondax) filters. Next, the beam shape is optimized and culminated by a pinhole spatial filter.
Control over incident and scattered polarization is realized by a set of two polarizers and two half-wave plates.
First, linear polarization is insured by a (Thorlabds) calcite polarizer. A monochromatic half-wave plate (HW0 in Fig.~\ref{fig:SM_setup}) is placed \textit{before} the polarizer and rotated for optimal incident intensity, correcting for the original laser polarization.
The beam then passes through a (NoiseBlock 90/10 Ondax) beam splitter (BS) with the experimental polarization orientation controlled by another monochromatic half-wave plate (HW1 in Fig.~\ref{fig:SM_setup}). 
The polarized beam then enters a microscope and focused on the sample by a 10X (Zeiss) objective.
Excitation powers of 7-30 mW were used (depending on temperature), measured just before the microscope.
The sample (see \ref{SM:Synthesis}) is placed inside a (TS1000 HiTemp) Linkam stage under Argon purge. 
Purging with a mono-atomic gas proved crucial, since under vacuum local laser heating would create defects, whisle nitrogen would dominate the Raman signal with its rotational excitations.  
A mono-atomic gas is also the best option for effective thermal coupling to the Linkam hot plate.
Inside the Linkam the magic happens, and laser light is scattered off the thermally excited crystal. 
The back-scattered light is collected by the same objective and passes back through HW1 and through the BS, where 90\% of the Rayleigh is eliminated and Raman signal transmitted.
Note that the beam's polarization at this point is not known, since we can not assume the scattered light, especially the Raman signal, scattered back into the same incident polarization, as is described in Fig.~\ref{fig:SM_setup} (\textbf{b}).
Indeed, there is no reason to assume the beam is still linearly polarized.  
We may, however, as always, describe the beam's polarization as the sum of two linear components, one parallel to the incident beam polarization, and the other perpendicular. 
After passing again through HW1, all light scattered in parallel to the incident beam polarization will now parallel the original polarization determined by the first polarizer. 
All light scattered perpendicular to the incident beam, will now be also polarized perpendicular to the first polarizer.  
After the BS, the scattered beam (orange line in Fig.~\ref{fig:SM_setup}) passes through an achromatic half-wave plate (HW2 in Fig.~\ref{fig:SM_setup}) which either leaves or rotates the signal polarization by 90\textsuperscript{0}. 
Another identical polarizer (designated "analyzer" in Fig.~\ref{fig:SM_setup}) filters the polarization of the signal. 
We thus separate the two scattered polarization components.
With HW2 at 0\textsuperscript{0}, only the parallel signal will make it past the analyzer, with HW2 at 45\textsuperscript{0}, only the perpendicular signal is transmitted.
The system achieved an extinction ratio $(I_{\perp}/I_{\parallel})$ between 1/500 and 1/100 depending on the orientation of HW1.
After the analyzer, the remaining Rayleigh component of the signal is attenuated by two (Ondax SureBlock ultra-narrowband) notch filters and the signal is directed into a one meter (Horiba FHR 1000) spectrometer and detected by a (Horiba Synapse) CCD.

\begin{figure*}[h]
\centering
\includegraphics[width=\textwidth]{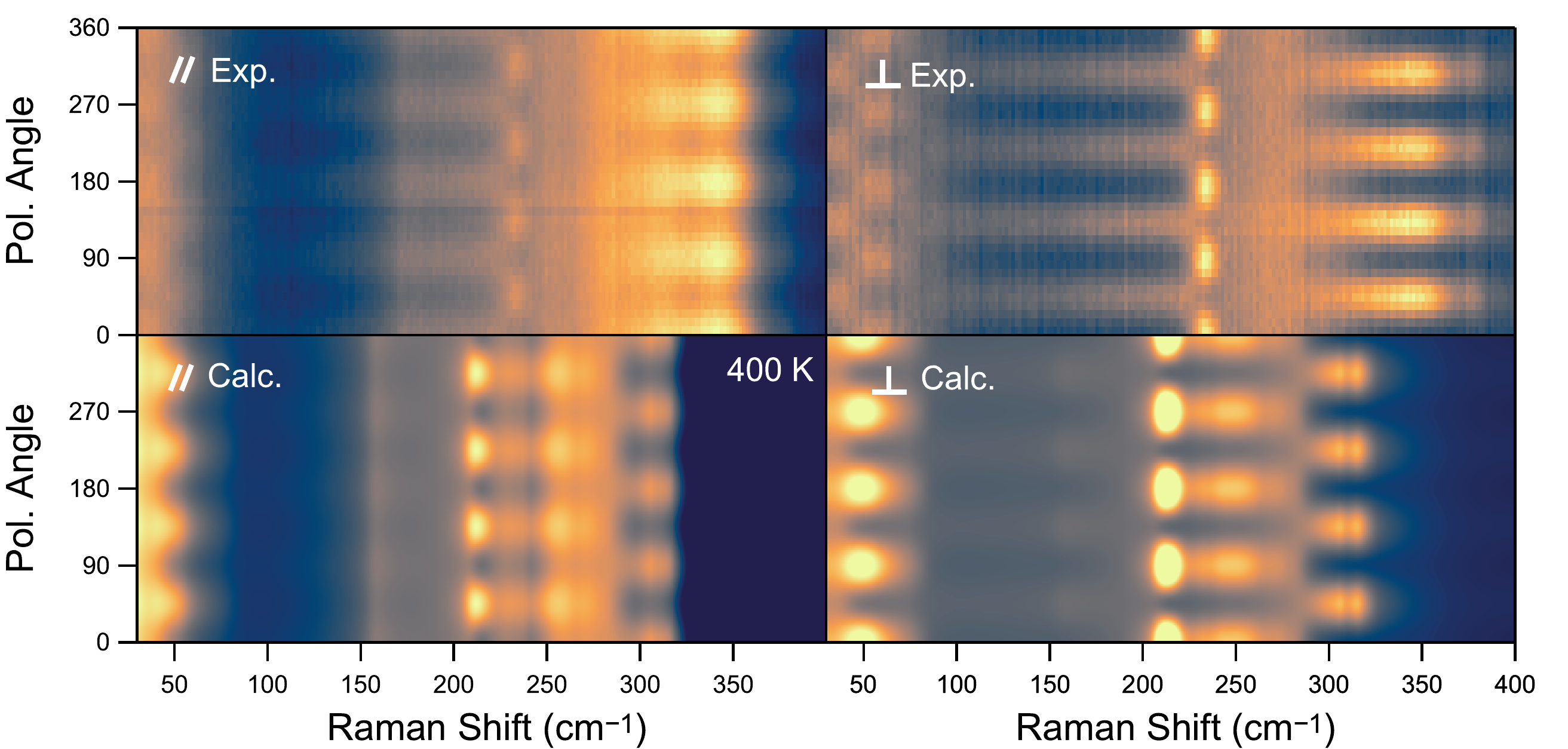}
\caption{NaCl Raman PO color map for measurement and TDEP simulation in 400~K. Top Row is for experimental measurement, bottom row for calculation. Left column is for parallel configuration and right is for perpendicular.}
\label{fig:NaClPO400K}
\end{figure*}
\begin{figure*}[h]
\centering
\includegraphics[width=\textwidth]{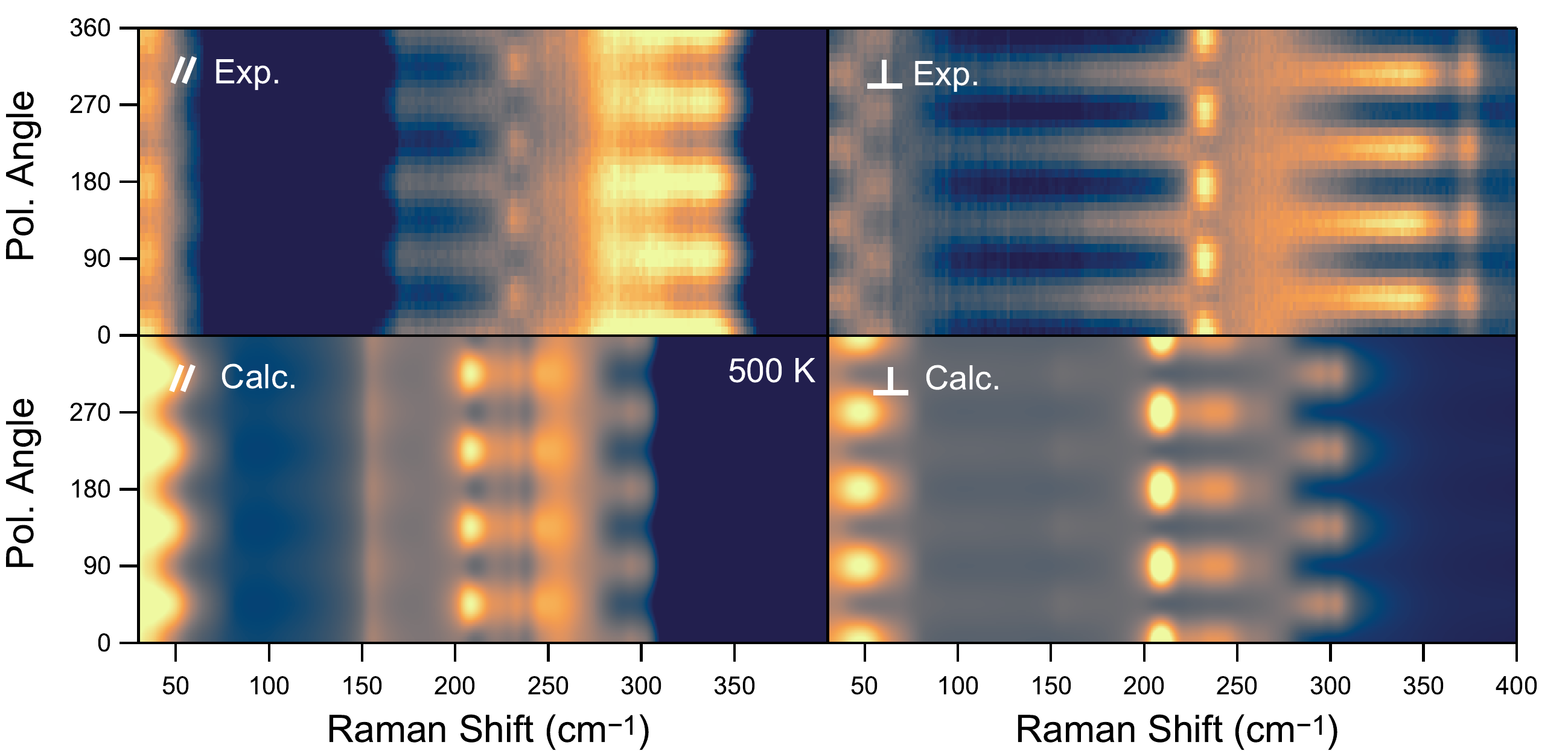}
\caption{NaCl Raman PO color map for measurement and TDEP simulation in 500~K. Top Row is for experimental measurement, bottom row for calculation. Left column is for parallel configuration and right is for perpendicular.}
\label{fig:NaClPO500K}
\end{figure*}
\begin{figure*}[h]
\centering
\includegraphics[width=\textwidth]{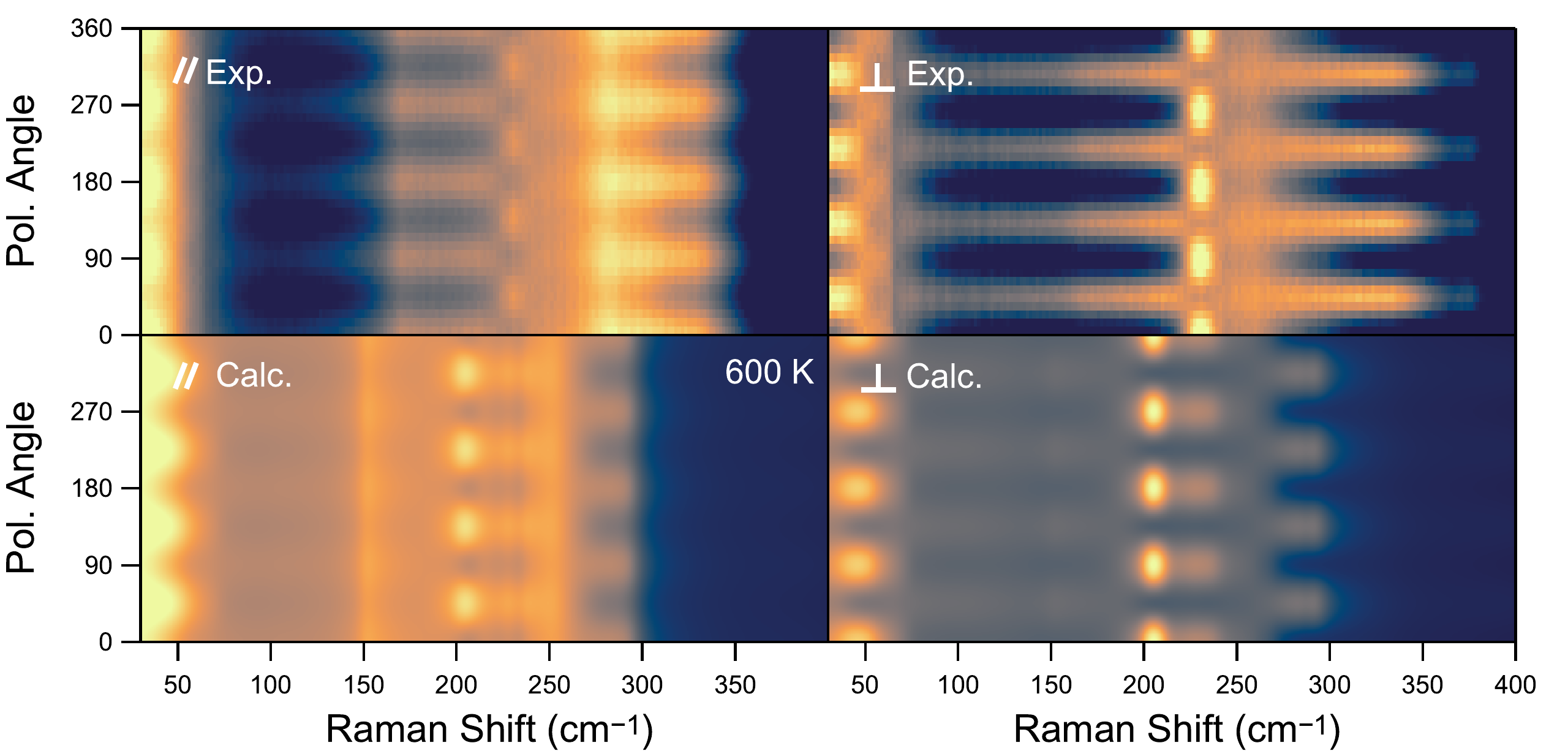}
\caption{NaCl Raman PO color map for measurement and TDEP simulation in 600~K. Top Row is for experimental measurement, bottom row for calculation. Left column is for parallel configuration and right is for perpendicular.}
\label{fig:NaClPO600K}
\end{figure*}
\begin{figure*}[h]
\centering
\includegraphics[width=\textwidth]{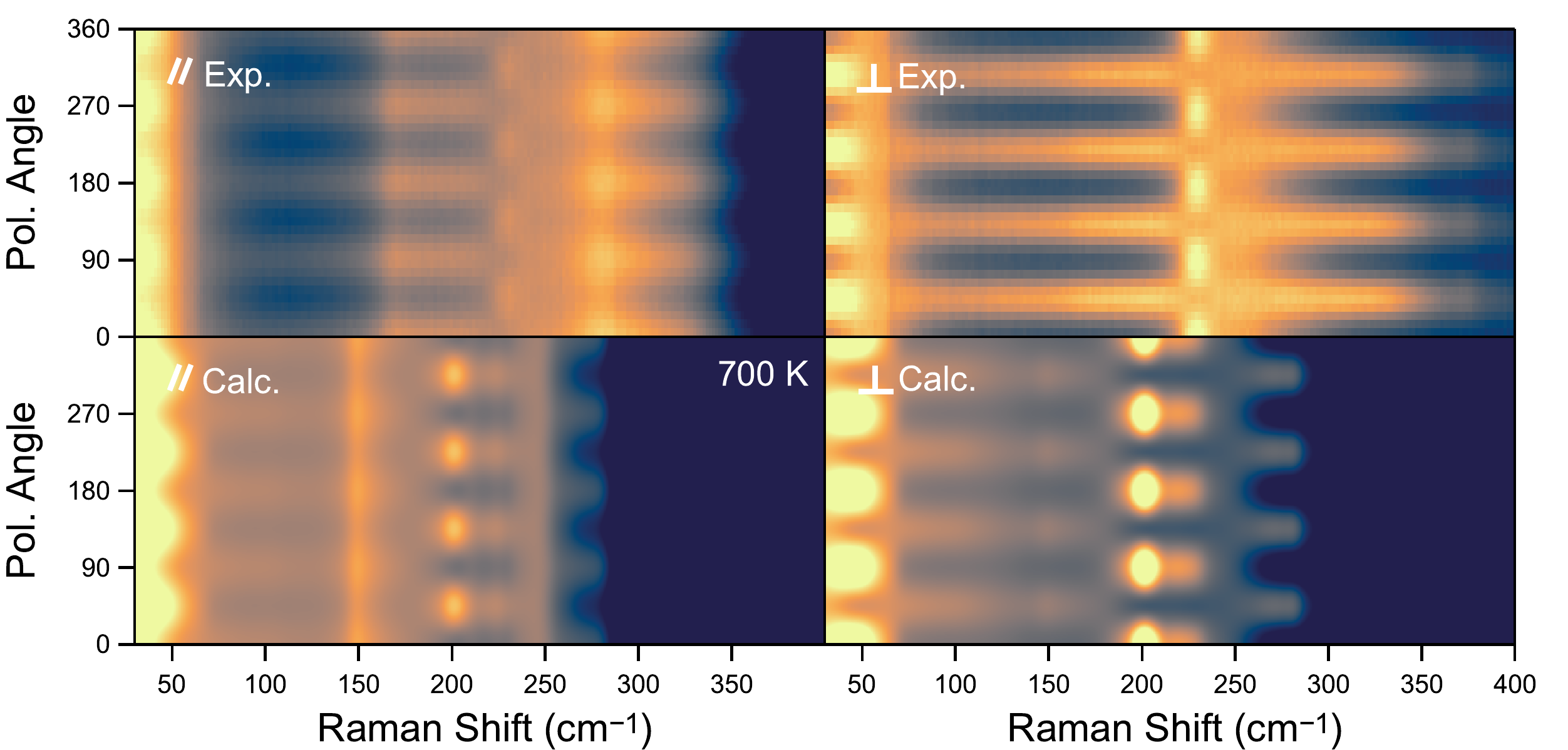}
\caption{NaCl Raman PO color map for measurement and TDEP simulation in 700~K. Top Row is for experimental measurement, bottom row for calculation. Left column is for parallel configuration and right is for perpendicular.}
\label{fig:NaClPO700K}
\end{figure*}

\vspace{1.5cm}
\section{Measurement and analysis} \label{SM:Measurement}

Here We give the detailed description of all data acquisition processing and present all PO maps for all measured temperatures.

To accumulate a Raman crystallography data-set, we rotate the polarization orientation of the incident laser beam $(\overrightarrow{\vec{E}}_i)$
in Fig.~\ref{fig:SM_setup} (\textbf{b}) in $10^0$ steps covering $\theta=0-360^0$ in a plane parallel to the NaCl (001) crystal face. 
At each $\theta$ step, a Raman spectrum is collected for both HW2 configurations (see $\hat{\vec{e}}_{\parallel}$ and $\hat{\vec{e}}_{\perp}$ in Fig.~\ref{fig:SM_setup}).

The signal of 2\textsuperscript{nd} order Raman is very weak, so long exposure times and multiple accumulations are necessary. 
For each measured spectrum, that is, for a given temperature, at a given incident polarization angle at given detection configuration, approximately 6 accumulations of 60 seconds were collected.

Some residual baseline persisted in all measured spectra, resembling a Lorentzian tail towards lower frequencies. This was cancelled in all unpolarized figures via standard baseline reduction. 
All PO maps are given in \cref{fig:NaClPO400K,fig:NaClPO500K,fig:NaClPO600K,fig:NaClPO700K}. These are presented with no augmentation other than a low-frequency cut-off, where the signal is governed by the notch filters and not the actual scattering. For each temperature, the corresponding TDEP calculation (see \ref{SM:green}) is also presented.

\vspace{1.5cm}
\section{Single Crystal Synthesis} \label{SM:Synthesis}

\begin{figure*}[h]
\centering
\includegraphics[width=0.5\textwidth]{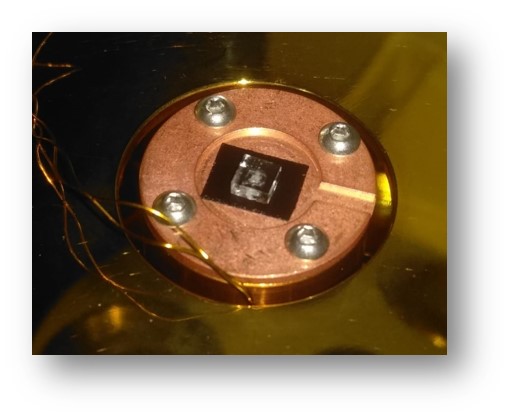}
\caption{One of the NaCl single crystals synthesized for the experiment..}
\label{fig:Single_Crystal}
\end{figure*}

The NaCl single crystal used in Raman measurements was home grown.
Double-deionized water (DDW) was heated slowly to around $40-50^0C$.
High purity $(\geq99\%)$ Sodium Chloride powder was added and stirred into the DDW until saturation.
The saturated solution was poured into another wide flask with a few fresh crystals for crystallization sites.
The flask was left covered with a filter paper (to avoid dust penetration and allow evaporation) in a chemical hood for several days until the crystals reached a size of $\sim$1cm.
Finally, the crystals were taken out of the dried flask and placed on a clean wipe to adsorb all residual water.
A picture of the crystal used is given in figure \ref{fig:Single_Crystal}.

\vspace{1.5cm}
\section{Factor group analysis for 2nd order Raman} \label{SM:Group_theory}

The full factor group analysis for the rock-salt structure has already been worked out by Burnstein~\cite{Burstein1965}, following methods established by Birman~\cite{Birman1962} and Elliot~\cite{Elliot1960}.
These ostensibly provide a complete tool-set to interpret 2\textsuperscript{nd} order Raman spectra by assigning each spectral feature to a known combination of vibrational modes.  
As mentioned in the main text, this treatment may prove useful for mode assignment of a known phononic dispersion relations to an observed spectrum. 
However, even in the high symmetry case of the cubic $O_h$ point group, selection rules for 2\textsuperscript{nd} order Raman are too relaxed to offer complete predictive power, even with reliable dispersion relations at hand. 

To demonstrate this, we perform the factor group analysis for for one high symmetry point in the Birllouin zone, specifically, for the $\mathrm{X'_5 (TO)},\:\mathrm{X'_5 (TA)}$ phonon pair in the reciprocal space X-point in a rock-salt structure.

The basic selection rule governing Raman scattering is~\cite{Yu2010}:
\begin{equation} \label{eq:selection_rule}
    \Gamma_f\subseteq\Gamma_{15}\otimes\Gamma_{15}
\end{equation}
with $\Gamma_f$ the representation of the excited phonon state, and $\Gamma_{15}$ the representation for the momentum operator in BSW notation (Raman scattering can be viewed as the product o two dipole transitions). The crux of the matter lies in the fact that in 2\textsuperscript{nd} order Raman this final phonon state is generally composed of two off $\Gamma$-point phonons, with its representation given by the Kronecker product of the two single-phonon states. For example:
\begin{align}
        \nonumber
       & \textrm{Quantum state:}\:\:
        \left|\mathrm{X}_{2phonon}\right\rangle =
        \left|\mathrm{X'_5 (TO),\:X'_5 (TA)}\right\rangle\:\to\: \\  \nonumber 
        \\ 
       & \textrm{Representation:}\:\:\:
        \mathrm{X}_{2phonon} =  
        \mathrm{X'_5 (TO)\otimes X'_5 (TA) = X_1 \oplus X_2 \oplus X_3 \oplus X_4}
\end{align}
with all the irreducible representations in the direct sum being Raman active according to Eq.~\eqref{eq:selection_rule}. This leads to the symmetry allowed Raman tensor form~\cite{Bilbao2006}:
\begin{equation}
    R_{\mathrm{X'}_5\times\mathrm{X'}_5} = R_{\mathrm{X}_1}+R_{\mathrm{X}_2}+R_{\mathrm{X}_3}+R_{\mathrm{X}_4}= \\
    \begin{pmatrix}
    a & d & 0 \\
    d & b & 0 \\
    0 & 0 & c 
    \end{pmatrix}
\end{equation}
which is a very general Raman tensor indeed. Now, consider that this tensor is responsible for the contribution to the scattered intensity inside some frequency interval $I(\Omega)\mathrm{d}\Omega$ from a single phonon combination.
In practice, equation $(11)$ allows a multitude of combinations to contribute to the same interval, such that accurately predicting the observed intensity, let alone the PO behavior of the signal, becomes intractable. 
The situation becomes even worse for lower symmetry structures, where more phonon combinations will be Raman active and have more general Raman tensors.

\vspace{1.5cm}
\section{Numerical determination of interaction tensors} \label{SM:Tensors}

Here we briefly reiterate the TDEP procedure for consistency of notation and to clarify the analogy between how dipole and polarizibility interactions are determined and how interatomic force constants are determined. The starting point is a set of displacements $\vec{u}$ and forces $\vec{f}$ from a set of $N_c$ supercells sampled from a canonical ensemble at temperature $T$. Although omitted in the notation, all interaction tensors depend on temperature as well as the volume (or more generally strain). The interatomic force constants are determined as follows: consider a supercell with $N_a$ atoms and forces and displacements given by the $3N_{a} \times 1$ vectors $\vec{u}$ and $\vec{f}$ (for clarity we will make note of the dimensions of the matrices in each step):

\begin{align}
    \underbrace{\vec{f}}_{3N_a \times 1} 
    & = - 
    \underbrace{\vec{\Phi}}_{3N_a \times 3 N_a} 
    \underbrace{\vec{u}}_{3N_a \times 1} 
\intertext{this can equivalently be written as}
    \underbrace{\vec{f}}_{3N_a \times 1} 
    & = - 
    \underbrace{(\vec{I} \otimes \vec{u}^T)}_{3N_a \times (3N_a)^2}
    \underbrace{\vec{\Phi}_v}_{ (3N_a)^2 \times 1}
\intertext{with the Kronecker product $\otimes$. To express the forces from $N_c$ supercells the matrices are stacked on top of each other:}
    \underbrace{
    \begin{pmatrix}
        \vec{f}_1 \\
        \vdots \\
        \vec{f}_{N_c}
    \end{pmatrix}
    }_{3N_a N_c \times 1} 
    & = - 
    \underbrace{
    \begin{pmatrix}
        \vec{I} \otimes \vec{u}_1^T \\
        \vdots \\
        \vec{I} \otimes \vec{u}_{N_c}^T \\
    \end{pmatrix}
    }_{3N_aN_c \times (3N_a)^2}
    \underbrace{\vec{\Phi}_v}_{ (3N_a)^2 \times 1}
\intertext{I can without any loss of generality express the second order force constants as a linear combination of $N_x$ irreducible components $\vec{x}^{\Phi\textrm{II}}$, where $N_x \le (3N_a)^2 $, via a matrix $\vec{C}^{\Phi\textrm{II}}$:}
    \underbrace{
    \begin{pmatrix}
        \vec{f}_1 \\
        \vdots \\
        \vec{f}_{N_c}
    \end{pmatrix}
    }_{3N_a N_c \times 1} 
    & = - 
    \underbrace{
    \underbrace{
    \begin{pmatrix}
        \vec{I} \otimes \vec{u}_1^T \\
        \vdots \\
        \vec{I} \otimes \vec{u}_{N_c}^T \\
    \end{pmatrix}
    }_{3N_aN_c \times (3N_a)^2}
    \underbrace{
    \vec{C}^{\Phi\textrm{II}}
    }_{(3N_a)^2 \times N_x}
    }_{=\vec{A}^{\Phi\textrm{II}}\,, 3N_a N_c \times N_x }
    \underbrace{\vec{x}^{\Phi\textrm{II}}}_{ N_x \times 1}
\end{align}
The set of symmetry relations and invariances determines $\vec{C}$ and $N_x$,~\cite{Leibfried1961,Maradudin1968,Born1998} and in general $N_x \ll (3N_a)^2$ which considerably simplifies the problem of numerically determining $\vec{x}$. In practice, only the matrix $\vec{A}$ is determined and stored. The nominally very large matrices that go into the construction of $\vec{A}$ are quite sparse and construction presents negligible computational cost. The second order force constants imply matrices that could possibly be treated naively, but the generalization to higher order quickly becomes intractable with dense matrix storage. The third order force constants, for example, comes down to
\begin{equation}
    \underbrace{
    \begin{pmatrix}
        \vec{f}_1 \\
        \vdots \\
        \vec{f}_{N_c}
    \end{pmatrix}
    }_{3N_a N_c \times 1} 
     = - 
    \underbrace{
    \underbrace{
    \begin{pmatrix}
        \vec{I} \otimes \vec{u}_1^T \otimes \vec{u}_1^T\\
        \vdots \\
        \vec{I} \otimes \vec{u}_{N_c}^T \otimes \vec{u}_{N_c}^T \\
    \end{pmatrix}
    }_{3N_aN_c \times (3N_a)^3}
    \underbrace{
    \vec{C}^{\Phi\textrm{III}}
    }_{(3N_a)^3 \times N_x}
    }_{=\vec{A}^{\Phi\textrm{III}}\,, 3N_a N_c \times N_x }
    \underbrace{\vec{x}^{\Phi\textrm{III}}}_{ N_x \times 1}    
\end{equation}
where the contracted matrix $\vec{A}$ is many orders of magnitude smaller than the matrices it is built from. The interatomic force constants are determined in succession:
\begin{align}
\label{eq:f2}
    \vec{A}^{\Phi\textrm{II}}
    \vec{x}^{\Phi\textrm{II}} & = \vec{f}
\\
\label{eq:f3}
    \vec{A}^{\Phi\textrm{III}}
    \vec{x}^{\Phi\textrm{III}} & = 
    \vec{f}-
    \vec{A}^{\Phi\textrm{II}}
    \vec{x}^{\Phi\textrm{II}}
\\    
\label{eq:f4}
    \vec{A}^{\Phi\textrm{IV}}
    \vec{x}^{\Phi\textrm{IV}} & = 
    \vec{f}
    -
    \vec{A}^{\Phi\textrm{II}}
    \vec{x}^{\Phi\textrm{II}}
    -
    \vec{A}^{\Phi\textrm{III}}
    \vec{x}^{\Phi\textrm{III}}
\end{align}
Where $\vec{f}$ denotes the reference forces to be reproduced. Equations \eqref{eq:f2}--\eqref{eq:f4} are overdetermined and solved as a least squares problem. This ensures that the baseline harmonic part becomes as large as possible, and that the higher order terms become smaller and smaller.

In a similar manner we search for a way to determine $M$ and $P$ (the expansion of the dipole moment and polarizability with respect to position) to all orders. The input is again a supercell, where in addition to forces we also have determined $\partial^2 U / \partial E^2$ as well as $\partial^2 U / \partial E \partial R$, that is the polarizability and Born charges. In a manner analogous to the force constants we determine the irreducible components of the interaction tensors and express the polarizability for a supercell as:
\begin{equation}
\begin{split}
    \underbrace{ \partial_E \partial_E U (\vec{u}) }_{9 \times 1}
    &
    =
    P_0 + 
    \underbrace{ \vec{I} \otimes \vec{u}^T }_{9 \times 27 N_a }
    \underbrace{ P }_{27 N_a \times 1 }
    +
    \underbrace{ \vec{I} \otimes \vec{u}^T \otimes \vec{u}^T }_{9 \times 9 (3N_a)^2 }
    \underbrace{ P }_{9 (3N_a)^2 \times 1 }
    +
    \ldots
\\    
    & = 
    P_0 + 
    \vec{A}^{P\textrm{I}}
    \vec{x}^{P\textrm{I}}
    +
    \vec{A}^{P\textrm{II}}
    \vec{x}^{P\textrm{II}}
    +
    \ldots
\end{split}    
\end{equation}
The terms in the polarizability expansion are determined slightly differently than the interatomic force constants. For a set of $N_c$ supercells I build the set of differences in polarizability between supercell $i$ and $j$:
\begin{align}
    \Delta^{0} P_{ij} & = 
    \partial_E \partial_E U(\vec{u}_i) -
    \partial_E \partial_E U(\vec{u}_j)
\intertext{and solve for $\vec{x}^{P\textrm{I}}$ from}
    \left(
    \vec{A}^{P\textrm{I}}_{i} -
    \vec{A}^{P\textrm{I}}_{j}
    \right)\vec{x}^{P\textrm{I}}
    & = \Delta^{0} P_{ij}
\intertext{where the set of all pairs of supercell $ij$ are used. The fluctuations from the first order are removed to construct}
    \Delta^{1} P_{ij} & = 
    \left(\partial_E \partial_E U(\vec{u}_i) - \vec{A}^{P\textrm{I}}_{i}\vec{x}^{P\textrm{I}}  \right)  -
    \left(
    \partial_E \partial_E U(\vec{u}_j) -
    \vec{A}^{P\textrm{I}}_{j}\vec{x}^{P\textrm{I}}
    \right)
\\
    \left(
    \vec{A}^{P\textrm{II}}_{i} -
    \vec{A}^{P\textrm{II}}_{j}
    \right)\vec{x}^{P\textrm{II}}
    & = \Delta^{1} P_{ij}
\end{align}
And so on for all orders considered. The baseline polarizability is then determined via
\begin{equation}
    P_0 + P^{\textrm{th}} = \left\langle 
    \partial_E \partial_E U -
    \vec{A}^{P\textrm{I}}
    \vec{x}^{P\textrm{I}} -
    \vec{A}^{P\textrm{II}}
    \vec{x}^{P\textrm{II}}
    - \ldots
    \right\rangle
\end{equation}
Where $P^{\textrm{th}}$ is the contribution to the susceptibility from the second order terms in the polarizability expansion~\cite{Cowley1964b}:
\begin{equation}
    P^{\textrm{th}} = \sum_{\lambda} P_{\lambda\lambda}(2n_{\lambda}+1)
\end{equation}
where the matrix elements are defined below.
For a perfectly harmonic system (or any harmonic sampling of phase space) the average of the first order term disappears.

The interaction parameters for the dipole moment are determined from the set of Born charges calculated in the supercell. We deliberately do not work with polarization directly. The quantized nature of polarization makes finding the polarization on the same branch for a large number of supercells tedious if not outright problematic. The starting point is again the idea of expressing the Born charges for a supercell in terms of the irreducible components:
\begin{equation}
\begin{split}
    \underbrace{ \partial_E \partial_R U }_{9N_a \times 1}(\vec{u})
    &
    =
    \underbrace{ \vec{M}^{\textrm{I}} }_{9N_a \times 1}
    +
    \underbrace{ \vec{I} \otimes \vec{u}^T }_{9N_a \times (9N_a)(3 N_a) }
    \underbrace{ M^{\textrm{II}} }_{(9N_a) (3N_a) \times 1 }
    +
    \underbrace{ \vec{I} \otimes \vec{u}^T \otimes \vec{u}^T }_{9N_a \times (9N_a)(3 N_a)^2 }
    \underbrace{ M^{\textrm{III}} }_{(9N_a) (3N_a)^2 \times 1 }
    +
    \ldots
    \\
    & = 
    \vec{A}^{\textrm{MI}}\vec{x}^{\textrm{MI}} + 
    \vec{A}^{\textrm{MII}}\vec{x}^{\textrm{MII}} + 
    \vec{A}^{\textrm{MIII}}\vec{x}^{\textrm{MIII}} + 
\end{split}    
\end{equation}
Where the terms are determined in succession:
\begin{align*}
    \vec{A}^{\textrm{MI}}\vec{x}^{\textrm{MI}} & = \partial_E \partial_R U
    \\
    \vec{A}^{\textrm{MII}}\vec{x}^{\textrm{MII}} & = \partial_E \partial_R U -\vec{A}^{\textrm{MI}}\vec{x}^{\textrm{MI}}
    \\
    \vec{A}^{\textrm{MIII}}\vec{x}^{\textrm{MIII}} & = \partial_E \partial_R U -\vec{A}^{\textrm{MI}}\vec{x}^{\textrm{MI}}
    -\vec{A}^{\textrm{MII}}\vec{x}^{\textrm{MII}}
\end{align*}

\vspace{1.5cm}
\section{Long-ranged electrostatics in polar materials}

Long-ranged interactions are not captured appropriately by the interatomic force constants and have to be treated with care. The formalism to deal with this in quasiharmonic calculations is well established, see e.g. \citet{Gonze1997}. For finite temperature calculations of interactions beyond pair interaction the formalism requires a slight adjustment.

We will first discuss the problem in quite general terms. Our starting point is a general long-ranged (polar) pair interaction $f(r)$ that we range-separate in some way:
\begin{equation}
    f(r) = 
    f^{\textrm{sr}}(r)+f^{\textrm{lr}}(r).
\end{equation}
If we set the range-separation such that the short-ranged $f^{\textrm{sr}}$ part fits snugly in the simulation cell, that part of the interaction is taken care of by the usual force constants and requires no additional consideration. The long-ranged tail $f^{\textrm{lr}}$ is what remains to be treated. This is done by calculating the Hessian of the long-ranged part:
\begin{equation}
    \Phi^{\textrm{lr}}_{ij\alpha\beta} = \left. \frac{\partial^2 f^{\textrm{lr}}(r)}{\partial u_{i}^\alpha \partial u_{j}^\beta} \right|_{r=r_{ij}}
\end{equation}
after which the forces from the long-ranged interactions can be calculated
\begin{equation}
    f_{i\alpha}^{\textrm{lr}} = -\sum_{j\beta} \Phi^{\textrm{lr}}_{ij\alpha\beta} u_{j\beta}
\end{equation}
these long-ranged forces are removed from the calculated forces, such that the procedure for determining the interatomic force constants becomes:
\begin{align}
\label{eq:ff2}
    \vec{A}^{\Phi\textrm{II}}
    \vec{x}^{\Phi\textrm{II}} & = \vec{f} - \vec{f}^{\textrm{lr}}
\\
\label{eq:ff3}
    \vec{A}^{\Phi\textrm{III}}
    \vec{x}^{\Phi\textrm{III}} & = 
    \vec{f}-\vec{f}^{\textrm{lr}}-
    \vec{A}^{\Phi\textrm{II}}
    \vec{x}^{\Phi\textrm{II}}
\\    
\label{eq:ff4}
    \vec{A}^{\Phi\textrm{IV}}
    \vec{x}^{\Phi\textrm{IV}} & = 
    \vec{f}- \vec{f}^{\textrm{lr}}
    -
    \vec{A}^{\Phi\textrm{II}}
    \vec{x}^{\Phi\textrm{II}}
    -
    \vec{A}^{\Phi\textrm{III}}
    \vec{x}^{\Phi\textrm{III}}
\end{align}
When determining the dynamical matrix, the long-ranged pair interactions are added back again:
\begin{equation}
    \vec{\Phi}(\vec{q}) = \vec{\Phi}^{\textrm{sr}}(\vec{q})+\vec{\Phi}^{\textrm{lr}}(\vec{q})
\end{equation}
This procedure is beneficial in several ways. By subtracting forces (instead of dynamical matrices as in \citet{Gonze1997}) numerical issues due to aliasing (finite size) errors are greatly reduced. More importantly, it also ensures there is no multiple counting: the long-ranged interactions are subtracted from interactions of all orders, ensuring the short-sightedness of higher order interactions. What we just described is the general procedure, for practical calculations one needs to assume an explicit form for the long-ranged interactions. We use the (now temperature-dependent) Born charges and dielectric constant, and the range-separatation is done using the standard Ewald technique. The resulting real-space force constants are then~\cite{Gonze1997}:
\begin{align}
\Phi^{\textrm{dd}}|^{\alpha\beta}_{ij} & = \sum_{\gamma\delta} Z_{i}^{\alpha\gamma}Z_{j}^{\beta\delta} \widetilde{\Phi}^{\gamma\delta}_{ij} \\
\widetilde{\Phi}^{\alpha\beta}_{ij} & =
\label{eq:SM_ddifc}
\frac{1}{4\pi\epsilon_0}
\frac{1}{\sqrt{\det \epsilon}}
\left(
	\frac
	{ \epsilon^{-1}_{\alpha\beta} }
	{|\Delta_{ij}|_{\epsilon}^{3}}
	-3 \frac
	{\Delta_{ij}^{\alpha}\Delta_{ij}^{\beta}}
	{|\Delta_{ij}|_{\epsilon}^{5}}
\right)
\end{align}
Here $\boldsymbol{\epsilon}$ is the dielectric tensor and $\Delta$ realspace distances using the dielectric tensor as a metric:
\begin{align}
\mathbf{r}_{ij} & = \mathbf{R}_{j}+\mathbf{\tau}_j-\mathbf{\tau}_i \\
\mathbf{\Delta}_{ij} & = \boldsymbol{\epsilon}^{-1}\mathbf{r}_{ij} \\
|\Delta_{ij}|_{\epsilon} & = \sqrt{\mathbf{\Delta}_{ij} \cdot \mathbf{r}_{ij}}
\end{align}
This poses an issue when calculating the dynamical matrix: the interactions die off as $~1/r^{3}$ which makes it necessary to extend the sum over lattice vectors to infinity. This is remedied with the usual Ewald technique:
\begin{equation}
\widetilde{\mathbf{\Phi}}_{ij} (\mathbf{q})=
\widetilde{\mathbf{\Phi}}^\textrm{r}+\widetilde{\mathbf{\Phi}}^\textrm{q}+\widetilde{\mathbf{\Phi}}^\textrm{c}
\end{equation}
Dividing the sum into a realspace part, a reciprocal part and a connecting part. The realspace part is given by
\begin{align}
\widetilde{\mathbf{\Phi}}^\textrm{r}_{ij} & =
-\frac{\Lambda^3}{4\pi\epsilon_0\sqrt{\det \epsilon}}
\sum_{\mathbf{R}}
\mathbf{H}(\Lambda \Delta_{ij},\Lambda|\Delta_{ij}|_{\epsilon})
e^{i \mathbf{q} \cdot \mathbf{R}} \\
\frac{\partial \widetilde{\mathbf{\Phi}}^\textrm{r}_{ij}}{\partial q_\alpha}  & =
-\frac{\Lambda^3}{4\pi\epsilon_0\sqrt{\det \epsilon}}
\sum_{\mathbf{R}}
iR_\alpha
\mathbf{H}(\Lambda \Delta_{ij},\Lambda|\Delta_{ij}|_{\epsilon})
e^{i \mathbf{q} \cdot \mathbf{R}}
\end{align}
where
\begin{equation}
H_{\alpha\beta}(\mathbf{x},y) = \frac{x_{\alpha}x_{\beta} }{y^2}
\left[
\frac{3\,\textrm{erfc}\,y}{y^3}
+
\frac{2 e^{-y^2}}{\sqrt{\pi}}\left(\frac{3}{y^2}+2 \right)
\right]
-\epsilon^{-1}_{\alpha\beta}
\left[
\frac{\textrm{erfc}\,y}{y^3} + \frac{2 e^{-y}}{\sqrt{\pi} y^2 }
\right]\,.
\end{equation}
In reciprocal space we have
\begin{align}
%
%
\label{eq:SM_philongrange}
	\widetilde{\Phi}_\textrm{q}|_{ij}
	&=
	\sum_{\mathbf{K}=\mathbf{G}+\mathbf{q}}
	\chi_{ij}(\mathbf{K},\Lambda)
	\left( \mathbf{K}\otimes\mathbf{K} \right) \\
%
%
	\frac{\partial}{\partial q_x} \widetilde{\Phi}_\textrm{q}|_{ij}
	&=
	\sum_{\mathbf{K}=\mathbf{G}+\mathbf{q}}
	\chi_{ij}(\mathbf{K},\Lambda)
	\left( \mathbf{K}\otimes\mathbf{K} \right)
	\left(
	i\tau^x_{ij} - \left[\sum_{\alpha} K_\alpha \epsilon_{\alpha x}
	\right]
	\left[ \frac{1}{\|\mathbf{K}\|_{\epsilon}}+\frac{1}{4\Lambda^2} \right]
	\right) +
	\chi_{ij}(\mathbf{K},\Lambda)
	\begin{pmatrix}
	2K_x & K_y & K_z \\
	K_y & 0 & 0 \\
	K_z & 0 & 0
	\end{pmatrix} \\
%
%
	\frac{\partial}{\partial q_y} \widetilde{\Phi}_\textrm{q}|_{ij}
	&=
	\sum_{\mathbf{K}=\mathbf{G}+\mathbf{q}}
	\chi_{ij}(\mathbf{K},\Lambda)
	\left( \mathbf{K}\otimes\mathbf{K} \right)
	\left(
	i\tau^y_{ij} - \left[\sum_{\alpha} K_\alpha \epsilon_{\alpha y}
	\right]
	\left[ \frac{1}{\|\mathbf{K}\|_{\epsilon}}+\frac{1}{4\Lambda^2} \right]
	\right) +
	\chi_{ij}(\mathbf{K},\Lambda)
	\begin{pmatrix}
	0 & K_x & 0 \\
	K_x & 2K_y & K_z \\
	0 & K_z & 0
	\end{pmatrix}\\
%
%
	\frac{\partial}{\partial q_z} \widetilde{\Phi}_\textrm{q}|_{ij}
	&=
	\sum_{\mathbf{K}=\mathbf{G}+\mathbf{q}}
	\chi_{ij}(\mathbf{K},\Lambda)
	\left( \mathbf{K}\otimes\mathbf{K} \right)
	\left(
	i\tau^z_{ij} - \left[\sum_{\alpha} K_\alpha \epsilon_{\alpha z}
	\right]
	\left[ \frac{1}{\|\mathbf{K}\|_{\epsilon}}+\frac{1}{4\Lambda^2} \right]
	\right) +
	\chi_{ij}(\mathbf{K},\Lambda)
	\begin{pmatrix}
	0 & 0 & K_x \\
	0 & 0 & K_y \\
	K_x & K_y & 2K_z
	\end{pmatrix}
\end{align}
where
\begin{align}
\chi_{ij}(\mathbf{K},\Lambda) &  =
\frac{1}{\Omega\epsilon_0}
\frac{
\exp\left(i \mathbf{q} \cdot \mathbf{\tau}_{ij} \right)
\exp\left( -\frac{\|\mathbf{K}\|_{\epsilon}}{4\Lambda^2}  \right)
}
{\|\mathbf{K}\|_{\epsilon}} \\
\|\mathbf{K}\|_{\epsilon} & =\sum_{\alpha\beta}\epsilon_{\alpha\beta}K_\alpha K_\beta
\end{align}
and finally a connecting part given by
\begin{equation}
\widetilde{\mathbf{\Phi}}_\textrm{c}|_{ij}=
\delta_{ij}
\frac{\Lambda^3}{3 \epsilon_0 \pi^{3/2} \sqrt{ \det \boldsymbol{\epsilon} } }
\widetilde{\boldsymbol{\epsilon}}
\end{equation}
This expression constitutes the dipole-dipole interactions at all distances. As explained in the beginning of the section we only want to subtract the long-ranged components from the forces. This separation is realized by choosing the Ewald parameter $\Lambda$ such that $\widetilde{\Phi}^r$ disappears at the supercell boundary, and keeping only the reciprocal space part of the Ewald sum. So, to summarize, the procedure is as follows:
\begin{itemize}
    \item Determine Ewald parameter $\Lambda$ such that the short-ranged interactions die off at the supercell boundary.
    \item Calculate $\vec{\Phi}^{\textrm{lr}}(\vec{q}=0)$ for the simulation supercell via Eq.~\eqref{eq:SM_philongrange}.
    \item Use $\vec{\Phi}^{\textrm{lr}}(\vec{q}=0)$ to calculate the forces $\vec{f}^{\textrm{lr}}$ originating from the long-ranged tails of the dipole-dipole interactions.
    \item Use equations \eqref{eq:ff2}--\eqref{eq:ff4} to determine interatomic force constants.
    \item When determining the dynamical matrix at any $\vec{q}$, add $\vec{\Phi}^{\textrm{lr}}(\vec{q})$ back again.
\end{itemize}
For reference we also specified the gradient of the dynamical matrix in case the group velocities are needed. 
To conclude this section we note that there is no principal difference between this approach and what is regularly used, but in practice the presented approach is easier to implement, suffers from less numerical artifacts and makes avoiding double-counting easier. 
The double-counting we refer to arises because if we only correct the dynamical matrices, some long-ranged polar interactions get picked up by the third and fourth order force constants. 
The non-analytical behavior of Eq.~\eqref{eq:SM_ddifc} is already included in the the long-ranged reciprocal space part and independent of Ewald coupling parameter.

\newpage
\section{Determining the Raman spectra}
\label{SM:green}

In this section we outline the procedure to determine Raman spectra of an anharmonic crystal. 
We will begin by sketching the procedure to obtain the phonon-phonon correlation functions in the presence of three-phonon anharmonicity via the equation of motion approach, since intermediate results are needed to obtain the polarizability-polarizability correlation functions.

If we express the displacements in terms of creation and annihilation operators we can identify the reciprocal space coefficients needed for the perturbative expansion, repeated here for clarity and consistency of notation. 
Start with the standard creation and annihilation operators for a phonon with momentum $\vec{q}$ and polarization $s$ (when there is no loss of clarity we use the compound index $\lambda$ to denote $\vec{q}s$):
\begin{align}
	a_{\vec{q}s} = & \frac{1}{\sqrt{2N\hbar}}
	\sum_{i\alpha} \epsilon_{\vec{q}s}^{i\alpha}
	\left( \sqrt{m_i \omega_{\vec{q}s}} u_{i\alpha}-i \frac{p_{i\alpha}}{ \sqrt{ m_i \omega_{\vec{q}s}} } \right) 
	e^{-i\mathbf{q}\cdot\mathbf{r}_i} 
\\
	a_{\vec{q}s}^\dagger = & \frac{1}{\sqrt{2N\hbar}}
	\sum_{i\alpha} \epsilon_{\vec{q}s}^{i\alpha\,\dagger} 
	\left( \sqrt{m_i \omega_{\vec{q}s}} u_{i\alpha} + i \frac{p_{i\alpha}}{ \sqrt{ m_i \omega_{\vec{q}s}} } \right) 
	e^{i\mathbf{q}\cdot\mathbf{r}_i}
\\
    A_{\vec{q}s} = a_{\vec{q}s} + a_{\bar{\vec{q}}s}^\dagger = &
\frac{2}{\sqrt{2N\hbar}}
	\sum_{i\alpha} \epsilon_{\vec{q}s}^{i\alpha}
	\sqrt{m_i \omega_{\vec{q}s}} u_{i\alpha}
	e^{-i\mathbf{q}\cdot\mathbf{r}_i} 
\\
    B_{\vec{q}s} = a_{\vec{q}s} - a_{\bar{\vec{q}}s}^\dagger = &
-\frac{2i}{\sqrt{2N\hbar}}
	\sum_{i\alpha} \epsilon_{\vec{q}s}^{i\alpha}
    \frac{p_{i\alpha}}{ \sqrt{ m_i \omega_{\vec{q}s}} }	
	e^{-i\mathbf{q}\cdot\mathbf{r}_i} 
\end{align}
Here $\epsilon$ are phonon eigenvectors ($\vec{\epsilon}_i \cdot \vec{\epsilon}_j = \delta_{ij}$), $m$ atomic masses, $\omega$ frequencies. The vector $\vec{r}$ is a lattice vector, and $u$ and $p$ are the position and momentum operators. We use  the notation $\bar{\vec{q}}=-\vec{q}$. 
I introduce scaled eigenvectors (to simplify notation) via
\begin{align}
	\upsilon_{\vec{q}s}^{i\alpha} = & \sqrt{ \frac{\hbar}{2 m_i \omega_{\vec{q}s} } } \epsilon_{\vec{q}s}^{i\alpha}
    %
%
\intertext{In this notation the matrix elements pertaining to anharmonicity become}
    \Phi_{\lambda\lambda'\lambda''} & = 
    \frac{1}{3!}
	\sum_{ijk\alpha\beta\gamma}
	\upsilon_{\vec{q}s}^{i\alpha}
	\upsilon_{\vec{q}'s'}^{j\beta}
	\upsilon_{\vec{q}''s''}^{k\gamma}
	\Phi_{ijk}^{\alpha\beta\gamma}
	e^{-i 
	(\vec{q}\cdot\vec{r}_i+ \vec{q}'\cdot\vec{r}_j+\vec{q}''\cdot\vec{r}_k)}
\\
    \Phi_{\lambda\lambda'\lambda''\lambda'''} & = 
    \frac{1}{4!}
	\sum_{ijkl\alpha\beta\gamma\delta}
	\upsilon_{\vec{q}s}^{i\alpha}
	\upsilon_{\vec{q}'s'}^{j\beta}
	\upsilon_{\vec{q}''s''}^{k\gamma}
	\upsilon_{\vec{q}'''s'''}^{l\delta}	
	\Phi_{ijkl}^{\alpha\beta\gamma\delta}
	e^{-i
	(\vec{q}\cdot\vec{r}_i +
	\vec{q}'\cdot\vec{r}_j +
	\vec{q}''\cdot\vec{r}_k +
	\vec{q}'''\cdot\vec{r}_l)
	} 
\intertext{so that the anharmonic part of the Hamiltonian becomes}
	H_A & = \Phi_{\lambda\lambda'\lambda''}A_{\lambda}A_{\lambda'}A_{\lambda''}+
	\Phi_{\lambda\lambda'\lambda''\lambda'''}A_{\lambda}A_{\lambda'}A_{\lambda''}A_{\lambda'''} + \ldots	
\end{align}
In an analogous way we define the coefficients of the expansion of the polarizability as
\begin{align}
    P^{\mu\nu}_{\lambda} & =
	\sum_{i\alpha}
	\upsilon_{\vec{q}s}^{i\alpha}
	P_{i}^{\mu\nu,\alpha}
	e^{-i\vec{q}\cdot\vec{r}_i}
	\Delta_{\vec{q}} 
	=
	\sum_{i\alpha}
	\upsilon_{\vec{\Gamma}s}^{i\alpha}
	P_{i}^{\mu\nu,\alpha}
	=
	P^{\mu\nu}(s)
\\
P^{\mu\nu}_{\lambda\lambda'} & =
    \frac{1}{2!}
	\sum_{ij\alpha\beta}
	\upsilon_{\vec{q}s}^{i\alpha}
	\upsilon_{\vec{q}'s'}^{j\beta}
	P_{ij}^{\mu\nu,\alpha\beta}
	e^{-i\vec{q}\cdot\vec{r}_i}
	e^{-i\vec{q}'\cdot\vec{r}_j}
	\Delta_{\vec{q}\vec{q}'} =
\\
	& =	
	\frac{1}{2!}
	\sum_{ij\alpha\beta}
	\upsilon_{\vec{q}s}^{i\alpha}
	(\upsilon_{\vec{q}s'}^{j\beta})^{\dagger}
	P_{ij}^{\mu\nu,\alpha\beta}
	e^{i\vec{q}\cdot(\vec{r}_j-\vec{r}_i)}
\\
	& =
	\frac{1}{2!}
	\sum_{ij\alpha\beta}
	(\upsilon_{\vec{q}s}^{i\alpha})^{\dagger}
	\upsilon_{\vec{q}s'}^{j\beta}
	P_{ij}^{\mu\nu,\alpha\beta}
	e^{-i\vec{q}\cdot(\vec{r}_j-\vec{r}_i)} = 
	P^{\mu\nu}(\vec{q},s,s') = 
	P^{\mu\nu}(\bar{\vec{q}},s,s')^\dagger =
	P^{\mu\nu}(\bar{\vec{q}},s',s)
\\
P^{\mu\nu}_{\lambda\lambda'\lambda''} & =
    \frac{1}{3!}
    \sum_{ijk\alpha\beta\gamma}
	\upsilon_{\vec{q}s}^{i\alpha}
	\upsilon_{\vec{q}'s'}^{j\beta}
	\upsilon_{\vec{q}'s'}^{k\gamma}
	P_{ijk}^{\mu\nu,\alpha\beta\gamma}
	e^{-i
	(\vec{q}\cdot\vec{r}_i+
	\vec{q}'\cdot\vec{r}_j+
	\vec{q}''\cdot\vec{r}_k)
	}
\intertext{and the dipole moment matrix elements become}
    M^{\mu}_{\lambda} & =
	\sum_{i\alpha}
	\upsilon_{\vec{\Gamma}s}^{i\alpha}
	M_{i}^{\mu,\alpha}
	=
	M^{\mu}(s)
\\
M^{\mu\nu}_{\lambda\lambda'} & =
    \frac{1}{2!}
	\sum_{ij\alpha\beta}
	(\upsilon_{\vec{q}s}^{i\alpha})^{\dagger}
	\upsilon_{\vec{q}s'}^{j\beta}
	M_{ij}^{\mu,\alpha\beta}
	e^{-i\vec{q}\cdot(\vec{r}_j-\vec{r}_i)} = 
	M^{\mu}(\vec{q},s,s') = 
	M^{\mu}(\bar{\vec{q}},s,s')^\dagger =
	M^{\mu}(\bar{\vec{q}},s',s)
\\
M^{\mu}_{\lambda\lambda'\lambda''} & =
    \frac{1}{3!}
    \sum_{ijk\alpha\beta\gamma}
	\upsilon_{\vec{q}s}^{i\alpha}
	\upsilon_{\vec{q}'s'}^{j\beta}
	\upsilon_{\vec{q}'s'}^{k\gamma}
	M_{ijk}^{\mu,\alpha\beta\gamma}
	e^{-i
	(\vec{q}\cdot\vec{r}_i+
	\vec{q}'\cdot\vec{r}_j+
	\vec{q}''\cdot\vec{r}_k)
	}
\intertext{Such that}
\label{eq:polarizabilityexpansion}
P^{\mu\nu} & = P^{\mu\nu}_0 + P^{\mu\nu}_{\lambda} A_{\lambda}
+ P^{\mu\nu}_{\lambda\lambda'} 
A_{\lambda}A_{\lambda'}
+ P^{\mu\nu}_{\lambda\lambda'\lambda''} 
A_{\lambda}A_{\lambda'}A_{\lambda''} + \ldots
\\
M^{\mu} & = M^{\mu}_0 + M^{\mu}_{\lambda} A_{\lambda}
+ M^{\mu}_{\lambda\lambda'} 
A_{\lambda}A_{\lambda'}
+ M^{\mu}_{\lambda\lambda'\lambda''} 
A_{\lambda}A_{\lambda'}A_{\lambda''} + \ldots
\end{align}
%

\subsection{Phonon Green's functions}

We briefly sketch the derivation of the phonon self-energy in the presence of third order anharmonicity since intermediate results will be used to express the Raman and infrared spectra. Starting from the retarded Green's function
\begin{equation}
    G^{X,Y} = -i\theta(t)\avg{[X(t),Y^\dagger(0)]}
\end{equation}
we express the equations of motion of the phonon-phonon Green's function via
\begin{subequations}
\begin{align}
\label{eq:dr0}
    \partial_t A(t) & = -i[H,A]
\\
	\frac{d}{d t} G^{AA}_{\lambda\lambda'} 
	& = - i \omega_{\lambda} G^{BA}_{\lambda\lambda'}
\\
	\frac{d}{d t} G^{BA}_{\lambda\lambda'}
	& = i 2 \delta_{ \lambda\lambda' }\delta( t )
	- i \omega_{\lambda} G^{AA}_{\lambda\lambda'}
	- i 6 \sum_{\mu_1 \mu_2} \Phi_{\lambda \mu_1 \mu_2} G^{AA,A}_{\mu_1 \mu_2,\lambda'}
\\
\label{eq:dr1}
	i \frac{d}{d t} G^{AA,A}_{\mu_1 \mu_2,\lambda'} & =
	\omega_{\mu_1} G^{BA,A}_{\mu_1 \mu_2,\lambda'} +
	\omega_{\mu_2} G^{AB,A}_{\mu_1 \mu_2,\lambda'}
\\
\label{eq:dr2}	
	i \frac{d}{d t} G^{AB,A}_{\mu_1 \mu_2,\lambda'} & =
	\omega_{\mu_1} G^{BB,A}_{\mu_1 \mu_2,\lambda'} +
	\omega_{\mu_2} G^{AA,A}_{\mu_1 \mu_2,\lambda'} +
	\sum_{\nu_1 \nu_2}
	6 \Phi_{\mu_2 \nu_1 \nu_2} 
	G^{AAA,A}_{\mu_1 \nu_1 \nu_2,\lambda'}
\\
\label{eq:dr3}	
	i \frac{d}{d t} G^{BA,A}_{\mu_1 \mu_2,\lambda'} & =
	\omega_{\mu_1} G^{AA,A}_{\mu_1 \mu_2,\lambda'} +
	\omega_{\mu_2} G^{BB,A}_{\mu_1 \mu_2,\lambda'} +
	\sum_{\nu_1 \nu_2}
	6 \Phi_{\nu_1 \nu_2 \mu_1} 
	G^{AAA,A}_{\nu_1 \nu_2 \mu_2,\lambda'}
\\
\label{eq:dr4}
	i \frac{d}{d t} G^{BB,A}_{\mu_1 \mu_2,\lambda'} & =
	\omega_{\mu_1} G^{AB,A}_{\mu_1 \mu_2,\lambda'} +
	\omega_{\mu_2} G^{BA,A}_{\mu_1 \mu_2,\lambda'}
\intertext{
the four-point Green's functions are decoupled~\cite{Semwal1972,Zubarev1960} via 
$\avg{abcd}\approx \avg{ab}\avg{cd} + \avg{ac} \avg{bd} + \avg{ad} \avg{bc}$ to yield
}
	\sum_{\nu_1 \nu_2}
	6 \Phi_{\mu_2 \nu_1 \nu_2} 
	G^{AAA,A}_{\mu_1 \nu_1 \nu_2,\lambda'}
	& =
	12 (2n_{\mu_1}+1) \sum_{\nu_1}\Phi_{\mu_1 \mu_2 \nu_1 } G^{AA}_{\nu_1 \lambda'}
\\
\label{eq:dr5}
	\sum_{\nu_1 \nu_2}
	6 \Phi_{\nu_1 \nu_2 \mu_1} 
	G^{AAA,A}_{\nu_1 \nu_2 \mu_2,\lambda'}
	& =
	12 (2n_{\mu_2}+1) \sum_{\nu_1}\Phi_{\mu_1 \mu_2 \nu_1 } G^{AA}_{\nu_1 \lambda'}
\end{align}
\end{subequations}
After Fourier transforming equations \eqref{eq:dr0}--\eqref{eq:dr5} to the frequency domain we get a solvable set of equations that recover the usual definition of three-phonon anharmonicity~\cite{Leibfried1961,Cowley1963,wallace1998thermodynamics,Semwal1972}:
\begin{equation}
    G_{\lambda\lambda'}(\Omega)^{-1} = G^0_{\lambda\lambda'}(\Omega)^{-1} + \Sigma_{\lambda\lambda'}
\end{equation}    
where
\begin{equation}
    \Sigma_{\lambda\lambda'}(Z) = -18 \sum_{\vec{q}_1\vec{q}_2 s_1 s_2}
    \Phi^{\lambda s_1s_2}_{\vec{q}\bar{\vec{q}}_1\bar{\vec{q}}_2}
    \Phi^{\lambda' s_1s_2}_{\bar{\vec{q}} \vec{q}_1 \vec{q}_2}
    S(s_1,s_2,Z)
\end{equation}
and
\begin{equation}
    S(s_a,s_b,Z) = 
	(n_{a}+n_{b}+1)
	\left[
	\frac{1}{(\omega_{a}+\omega_{b}-Z)_p}-
	\frac{1}{(\omega_{a}+\omega_{b}+Z)_p}
	\right]
	+
	(n_{a}-n_{b})
	\left[
	\frac{1}{(\omega_{b}-\omega_{a}+Z)_p}-
	\frac{1}{(\omega_{b}-\omega_{a}-Z)_p}
	\right]
\end{equation}
Naturally if one considers the four-phonon interactions you recover the standard results, but the information we need right now is the intermediate result for the three-point Green's function expressed in terms of two-point Green's functions and three-phonon matrix elements:
\begin{equation}
    \label{eq:threepointGF}
    G_{\lambda\lambda',\lambda''} =
    6 \sum_{s} \Phi_{\bar{\lambda}\bar{\lambda}'s} 
    S(\lambda,\lambda')
    G_{ A_{s} A_{\lambda''} }
\end{equation}
\subsection{Raman spectra}
If we start from the polarizability-polarizability correlation function and insert the expansion of the polarizability in terms of phonon coordinates, Eq.~\eqref{eq:polarizabilityexpansion}, we get
\begin{equation}
\begin{split}
    \avg{P(t)P(0)} = i\overset{>}{G}(P,P) = &
    \vec{P}_0 \otimes \vec{P}_0 + 
    \vec{P}_{\lambda} \otimes \vec{P}_{\lambda'} i\overset{>}{G}(A_{\lambda},A_{\lambda'})
+ \\
    + &
    \vec{P}_{\lambda\lambda'}
    \otimes
    \vec{P}_{\lambda''\lambda'''}
    i\overset{>}{G}(A_{\lambda}A_{\lambda'},A_{\lambda''}A_{\lambda'''})
+ \\    
     + &
    \vec{P}_{\lambda} \otimes \vec{P}_{\lambda'\lambda''}
    i\overset{>}{G}(A_{\lambda},A_{\lambda'}A_{\lambda''})
+
    \vec{P}_{\lambda\lambda'}
    \otimes
    \vec{P}_{\lambda''}
    i\overset{>}{G}(A_{\lambda}A_{\lambda'},A_{\lambda''})
+ \\
    + &
\vec{P}_{\lambda} \otimes \vec{P}_{\lambda'\lambda''\lambda'''}
    i\overset{>}{G}(A_{\lambda},A_{\lambda'}A_{\lambda''}A_{\lambda'''})
+
    \vec{P}_{\lambda\lambda'\lambda''}
    \otimes
    \vec{P}_{\lambda'''}
    i\overset{>}{G}(A_{\lambda}A_{\lambda'}A_{\lambda''},A_{\lambda'''})
\end{split}
\end{equation}
where we have omitted terms of $A^4$ or higher. We will deal with the terms in order. The constant term does not contribute to the Raman spectrum. The first term that gives a contribution is
\begin{equation}
\label{eq:SMramanI}
    I^{\textrm{I}} =
    \vec{P}_{\lambda} \otimes \vec{P}_{\lambda'}
    \int
    i \overset{>}{G}(A_{\lambda}(t),A_{\lambda'}(0))
    e^{-i\Omega t} dt
    =
    \vec{P}_{\lambda} \otimes \vec{P}_{\lambda'}
    (n(\Omega)+1)
    J_{\lambda\lambda'}(\Omega)
\end{equation}
where $J$ is the phonon spectral function
\begin{equation}
	J(\Omega) = 
	- \frac{1}{\pi} \Im\left\{ G(\Omega) \right\} =	\frac{i}{(n(\Omega,T)+1)}\overset{>}{G}(\Omega)
\end{equation}
We identify Eq.~\eqref{eq:SMramanI} as the first order Raman spectra. 
%
The next contribution comes from
\begin{equation} \label{eq:2nd_ord_Raman}
    I^{\textrm{II}} =
    \vec{P}_{\lambda\lambda'} \otimes \vec{P}_{\lambda''\lambda'''}
    \int
    \overset{>}{G}(A_{\lambda}(t)A_{\lambda'}(t),A_{\lambda''}(0)A_{\lambda'''}(0))
    e^{-i\Omega t} dt
\end{equation}
The first step is to decouple the four-point Green's function~\cite{Zubarev1960}:
\begin{equation}
\begin{split}
    \overset{>}{G}(A_{\lambda}(t)A_{\lambda'}(t),A_{\lambda''}(0)A_{\lambda'''}(0))
  & \approx  
    \overset{>}{G}(A_{\lambda}(t),A_{\lambda'''}(0))
    \overset{>}{G}(A_{\lambda'}(t),A_{\lambda''}(0))
    +
\\  & +      
    \overset{>}{G}(A_{\lambda}(t),A_{\lambda''}(0))
    \overset{>}{G}(A_{\lambda'}(t),A_{\lambda'''}(0))
    +
\\  & +
    \underbrace{\overset{>}{G}(A_{\lambda}(t),A_{\lambda'}(t))
    \overset{>}{G}(A_{\lambda''}(0),A_{\lambda'''}(0))}_{= \textrm{constant}}
\end{split}    
\end{equation}
and note that the Fourier transform of a product becomes a convolution:
\begin{equation}
\begin{split}
    I^{\textrm{II}} & =
    \vec{P}_{\lambda\lambda'} \otimes \vec{P}_{\lambda''\lambda'''}
    \frac{1}{\pi^2}
    \bigg[    
    \int
    (n(\Omega')+1)
    J_{\lambda\lambda''}(\Omega')
    (n(\Omega-\Omega')+1) J_{\lambda'\lambda'''}(\Omega-\Omega')
    d\Omega'
    +
\\
    & +
    \int
    (n(\Omega')+1)
    J_{\lambda\lambda'''}(\Omega')
    (n(\Omega-\Omega')+1) J_{\lambda'\lambda''}(\Omega-\Omega')
    d\Omega'
    \bigg]
    =
\\    
& = 2\Re\left\{ 
\vec{P}_{\lambda\lambda'} \otimes \vec{P}_{\lambda''\lambda'''}
 \right\}
     \int
    (n(\Omega')+1)
    J_{\lambda\lambda''}(\Omega')
    (n(\Omega-\Omega')+1) J_{\lambda'\lambda'''}(\Omega-\Omega')
    d\Omega' 
\end{split}    
\end{equation}
where we relabelled the summation indices in the second term. 
This term is the second-order Raman spectra, slightly more general than it is usually presented. 
If we assume a diagonal self-energy (and consequently a diagonal spectral function) and insert the non-interacting spectral function $J^0_{\lambda} = \delta(\Omega-\omega_{\lambda}) - \delta(\Omega+\omega_{\lambda})$ the integral becomes
\begin{equation}
\begin{split}
    & \int
    n(-\Omega')
    \left[
        \delta(\Omega'-\omega_a)
        -
        \delta(\Omega'+\omega_a)
    \right]
    n(\Omega'-\Omega)
    \left[
        \delta(\Omega-\Omega'-\omega_b)
        -
        \delta(\Omega-\Omega'+\omega_b)
    \right]
    d\Omega' =
\\
    = &
    n(-\omega_a)
    n(\omega_a-\Omega)
    \left[
        \delta(\Omega-\omega_a-\omega_b)
        -
        \delta(\Omega-\omega_a+\omega_b)
    \right] -
\\    
    & 
    n(\omega_a)
    n(-\omega_a-\Omega)
    \left[
        \delta(\Omega+\omega_a-\omega_b)
        -
        \delta(\Omega+\omega_a+\omega_b)
    \right] =
\\
    = &
    \left[n(\Omega)+1\right]
    \left[
    n(\omega_a)+n(\omega_b)+1
    \right]
    \left[
    \delta(\Omega+\omega_a+\omega_b)
    -
    \delta(\Omega-\omega_a-\omega_b)
    \right]
    +
\\
    & +
    \left[n(\Omega)+1\right]
    \left[
    n(\omega_a)-n(\omega_b)
    \right]
    \left[
    \delta(\Omega-\omega_a+\omega_b)
    -
    \delta(\Omega+\omega_a-\omega_b)
    \right] =
\\
    = & -\frac{n(\Omega)+1}{\pi} \Im\left\{ S(\omega_a,\omega_b,\Omega) \right\}
\end{split}
\end{equation}
where we made use of $n(a+b) = (1+n(a)+n(b))/n(a)n(b)$ as well as $n(-a)=-n(a)-1$. 
With this we recover the expression given by \citet{Cowley1964b} as the limit of weak anharmonicity. 
In this study, however, it was found necessary to keep the interacting spectral function when calculating the Raman spectra. 
It is intuitive: the second order Raman is proportional to the two-phonon DOS, and if the phonons are broadened then the two-phonon DOS should also be broadened.

Next we have a term given by
\begin{equation}
    I^{\textrm{III}} =
    \vec{P}_{\lambda} \otimes \vec{P}_{\lambda'\lambda''}
    G^{>}(A_{\lambda},A_{\lambda'}A_{\lambda''})
+
    \vec{P}_{\lambda\lambda'}
    \otimes
    \vec{P}_{\lambda''}
    G^{>}(A_{\lambda}A_{\lambda'},A_{\lambda''})
\end{equation}
where we make use of the intermediate result for the three-point Green's functions in Eq.~\eqref{eq:threepointGF} and arrive at
\begin{equation}
I^{\textrm{III}} =
-6
(n(\Omega)+1)
\sum_{\mu}
\left(
\vec{P}_{\lambda''\lambda'}
\otimes
\vec{P}_{\lambda}
\Phi_{\bar{\lambda}''\bar{\lambda}'\mu}
+
\vec{P}_{\lambda} \otimes \vec{P}_{\lambda'\lambda''}
\Phi_{\bar{\lambda}'\bar{\lambda}''\mu}
\right)
\Im\left\{ 
S_{\lambda''\lambda'}
\right\}
J_{\mu \lambda}(\Omega)
\end{equation}
where, if we assume a diagonal self-energy, we recover previous results.~\cite{Cowley1964b} 
It is worth noting is that -- as far as we have discovered -- the last two terms are in practice several orders of magnitude smaller than the usual first and second order terms. 
In term III we have a product between the two-phonon DOS and the one-phonon spectral function, so to have any sizeable contribution these need to peak simultaneously. 
The final term
\begin{equation}
    I^{\textrm{IV}} =
\vec{P}_{\lambda} \otimes \vec{P}_{\lambda'\lambda''\lambda'''}
    G(A_{\lambda},A_{\lambda'}A_{\lambda''}A_{\lambda'''})
+
    \vec{P}_{\lambda\lambda'\lambda''}
    \otimes
    \vec{P}_{\lambda'''}
    G(A_{\lambda}A_{\lambda'}A_{\lambda''},A_{\lambda'''})    
\end{equation}
we approach with the same decoupling procedure as before:
\begin{equation}
\begin{split}
    G^{>}(A_{\lambda}(t),A_{\lambda'}(0),A_{\lambda''}(0)A_{\lambda'''}(0))
  & \approx  
    G^{>}(A_{\lambda}(t),A_{\lambda'''}(0))
    G^{>}(A_{\lambda'}(0),A_{\lambda''}(0))
    +
\\  & +      
    G^{>}(A_{\lambda}(t),A_{\lambda''}(0))
    G^{>}(A_{\lambda'}(0),A_{\lambda'''}(0))
    +
\\  & +
    G^{>}(A_{\lambda}(t),A_{\lambda'}(0))
    G^{>}(A_{\lambda''}(0),A_{\lambda'''}(0)) =
\\
    & = 3 G^{>}(A_{\lambda}(t),A_{\lambda'''}(0))\delta_{\lambda'\lambda''}(2n_{\lambda'}+1)
\end{split}    
\end{equation}
where the factor 3 comes from relabelling indices. 
We also used the equal-time thermal average of the phonon operators. This gives
\begin{equation}
    I^{\textrm{IV}} =
    3
    (n(\Omega)+1)
    \left(
    \vec{P}_{\lambda} \otimes \vec{P}_{\lambda'\lambda'\lambda''}
    +
    \vec{P}_{\lambda''\lambda'\lambda'}
    \otimes
    \vec{P}_{\lambda}
    \right)
    (2n_{\lambda'}+1)
    J_{\lambda\lambda''}
\end{equation}
What we derived in the section above holds for any operator that is expressed as a constant multiplied with the phonon operator, i.e. a quantity that only depends on the position of the atoms (it's frequency dependence can be ignored, or at least the cross-terms between the frequency dependence and the position of atoms). 
The derivation of the infrared absorption spectrum is identical, one only needs to replace the matrix elements. 
If the real part of the susceptibility is of interest it is conveniently calculated via a Kramers-Kronig transformation of the imaginary part.

In the main text we have defined compound matrix elements for the polarizability to simplify the notation, they are given by:
\begin{align}
	P^{(\textrm{I})}_{\mu\nu,\xi\rho}(s_1) & = P^{\mu\nu}(s_1) P^{\xi\rho}(s_1)^\dagger	
\\
	P^{(\textrm{II})}_{\mu\nu,\xi\rho}(\vec{q},s_1,s_2) & =
	P^{\mu\nu} (\vec{q},s_1,s_2)
	P^{\xi\rho} (\vec{q},s_1,s_2)^\dagger	
\\
	P^{(\textrm{III})}_{\mu\nu,\xi\rho}(\vec{q},s_1,s_2,s_3) & =
	\Phi_{\vec{q} \bar{\vec{q}} \vec{\Gamma}}^{s_1 s_2 s_3}
	P^{\mu\nu}( \vec{q},s_1,s_2 )	
	P^{\xi\rho}(s_3)
	+
	\Phi_{\vec{q} \bar{\vec{q}} \vec{\Gamma}}^{s_1 s_2 s_3}
	P^{\xi\rho}( \vec{q},s_1,s_2 )	
	P^{\mu\nu}(s_3)
\\
	P^{(\textrm{IV})}_{\mu\nu,\xi\rho}(\vec{q},s_1,s_2) & =
	P^{\mu\nu}(\vec{\Gamma},\bar{\vec{q}},\vec{q},s_1,s_2,s_2)
	P^{\xi\rho}(s_1)
	+
	P^{\xi\rho}(\vec{\Gamma},\bar{\vec{q}},\vec{q},s_1,s_2,s_2)
	P^{\mu\nu}(s_1)
\intertext{In exactly the same way I can define the following dipole matrix elements:}
	\label{eq:ircomp1}
	M^{(I)}_{\alpha\beta}(s_1) & = M^{\alpha}_{s_1} (M^{\beta}_{s_1})^\dagger
	\\
	\label{eq:ircomp2}
	M^{(II)}_{\alpha\beta}(s_1,s_2,\vec{q}) & =
	M^\alpha (^{s_1 s_2}_{\vec{q} })
	M^\beta (^{s_1 s_2}_{\vec{q} })^\dagger
	\\
	\label{eq:ircomp3}
	M^{(III)}_{\alpha\beta}(s_1,s_2,s_3,\vec{q}) & =	
	\Phi_{\vec{q} \bar{\vec{q}} \vec{0}}^{s_1 s_2 s_3}
	M^\beta(^{s_1 s_2}_{\vec{q}} )	
	M^\alpha_{s_3}
	+
	\Phi_{\vec{q} \bar{\vec{q}} \vec{0}}^{s_1 s_2 s_3}
	M^\alpha(^{s_1 s_2}_{\vec{q}} )	
	M^\beta_{s_3}
	\\
	\label{eq:ircomp4}
	M^{(IV)}_{\alpha\beta}(s_1,s_2,\vec{q}) & =		
	M^\alpha(^{s_1 s_2 s_2}_{ \vec{0}\bar{\vec{q}}\vec{q} })
	M^\beta_{s_1}
	+
	M^\beta(^{s_1 s_2 s_2}_{ \vec{0}\bar{\vec{q}}\vec{q} })
	M^\alpha_{s_1}
\end{align}
Here I have explicitly used momentum conservation in the matrix elements. There is only a single q-vector that remains, which means that the computational cost of determining the spectras (given interaction tensors) are modest, at most a single sum over the Brillouin zone.

\vspace{1.5cm}
\section{Details of calculations}  \label{SM:details}

\subsection{Self-consistency procedure}

The interatomic force constants are determined self-consistently as explained in \citet{Shulumba2016b}, but it is worth explaining the practical procedure. 
The basic idea is that a set of force constants gives us normal modes that can be thermally populated, generating thermally excited structures. 
Thermally excited structures can, with the help of equation \eqref{eq:ff2} yield a new set of force constants. 
The procedure is terminated when the input and output force constants show no significant difference.

The self-consistency procedure is implemented via a geometric series:
\begin{itemize}
    \item Create an initial guess for the second order force constants (see \citet{Shulumba2016b})  and generate 1 thermally excited supercell.
    \item Generate new force constants and produce 2 thermally excited supercells.
    \item Use both the one old and two new supercells to generate a new set of force constants, and 4 new supercells.
    \item Use 2+4 supercells to generate forceconstants and 8 supercells.
    \item Use 4+8 supercells to generate forceconstants and 16 supercells.
    \item Repeat until convergence.
\end{itemize}
This procedure has the benefit of introducing some mixing (since you use supercells generated from different force constants) smoothing convergence. 
In addition, the prescribed procedure leaves only 25\% of calculations as waste. 
A straight implementation (generate N supercells each iteration, repeat until convergence) can waste a much larger fraction of calculations.

For this study all calculated spectra converged after 7 iterations, using 128+64 configurations. 
The calculations were done on a grid of volumes and temperatures, and a few points were tested with one more iteration where no appreciable difference in calculated quantities could be seen.

\begin{figure*}[hbt]
     \centering
     \includegraphics[width=\linewidth]{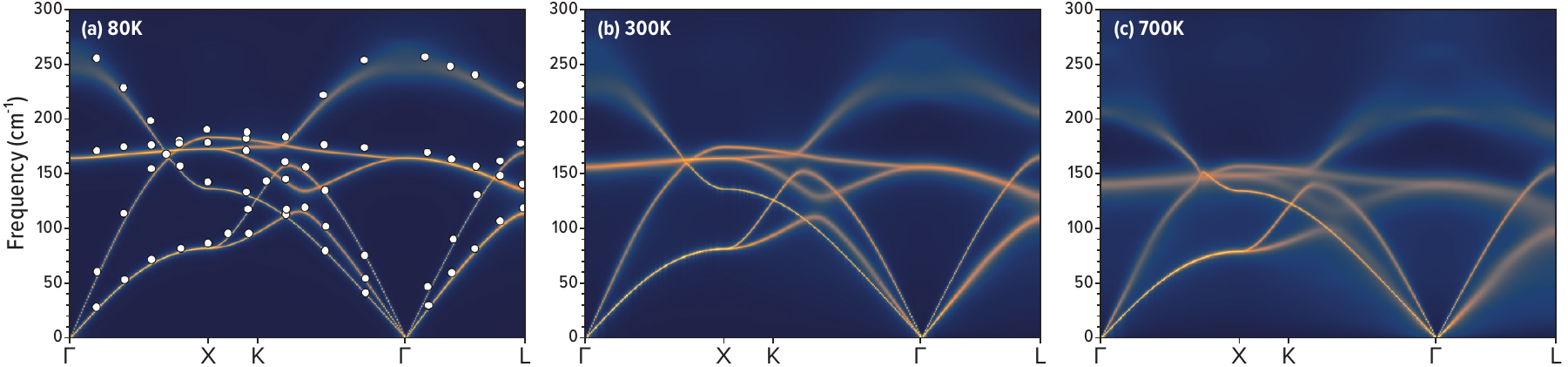}
     \caption{Phonon spectral functions for NaCl at different temperatures calculated using the TDEP method. The white circles in panel (a) are experimental neutron measurements from \citet{Raunio1969a}. We show an overall good agreement with experiment bar a slight uniform underestimation of frequencies.}
    \label{fig:SMspectralfunction}
\end{figure*}

\subsection{Free energy minimization}

To determine the equilibrium volume at any given temperature we did calculations on a grid of volumes and temperatures. The grid is specified in Table \ref{table:SM_simulationgrid}.
\begin{table*}
\begin{ruledtabular}
\begin{tabular}{r|llllll}
Temperature (K) & \multicolumn{6}{c}{Lattice constants (\AA)} \\
\hline
0 & 5.233848 & 5.324865 & 5.419088 & 5.510051 & 5.603468 & 5.717206 \\ 
150 & 5.255144 & 5.346531 & 5.441138 & 5.532472 & 5.626268 & 5.740469 \\ 
300 & 5.282018 & 5.373873 & 5.468963 & 5.560764 & 5.655040 & 5.769825 \\ 
500 & 5.326527 & 5.419155 & 5.515047 & 5.607621 & 5.702692 & 5.818444 \\ 
700 & 5.380951 & 5.474526 & 5.571398 & 5.664918 & 5.760960 & 5.877895 \\ 
1000 & 5.481180 & 5.576498 & 5.675174 & 5.770436 & 5.868267 & 5.987380 \\
\end{tabular}
\end{ruledtabular}
\caption{\label{table:SM_simulationgrid}
List of temperatures and lattice parameters used in the simulations.}
\end{table*}
Once simulations are converged the free energy was determined via
\begin{equation}
    F = U_0 + F_{\textrm{ph}} + \Delta F^{3\textrm{ph}} + \Delta F^{4\textrm{ph}}
\end{equation}
where $F_{\textrm{ph}}$ is the usual phonon free energy and $U_0$ the renormalized baseline free energy given by
\begin{equation}
    U_0 = \avg{U -
    \frac{1}{2!}\sum_{\substack{ ij\\ \alpha\beta } }\overset{\textrm{lr}}{\Phi}_{ij}^{\alpha\beta}
u_i^\alpha u_j^\beta - 
    \frac{1}{2!}\sum_{\substack{ ij\\ \alpha\beta } }\Phi_{ij}^{\alpha\beta}
u_i^\alpha u_j^\beta - 
 \frac{1}{3!}
\sum_{\substack{ijk\\ \alpha\beta\gamma}}\Phi_{ijk}^{\alpha\beta\gamma}
u_i^\alpha u_j^\beta u_k^\gamma - 
\frac{1}{4!}
	\sum_{\substack{
	ijkl\\
	\alpha\beta\gamma\delta
	}}
\Phi_{ijkl}^{\alpha\beta\gamma\delta}
u_i^\alpha u_j^\beta u_k^\gamma u_l^\delta 
    }
\end{equation}
Here it is important to note that we have to subtract the long-ranged polar interactions to avoid double-counting them. The explicit anharmonic contributions are given via~\cite{Leibfried1961,Cowley1963,wallace1998thermodynamics}
\begin{equation}\label{eq:deltaF3}
	\Delta F^{3\textrm{ph}} =
	-6
	\sum_{\lambda\lambda'\lambda''}
	\left|
		\Phi_{\lambda\lambda'\lambda''}
	\right|^2 
	\left(
	\frac{3n_{\lambda} n_{\lambda'} + 3n_{\lambda} + 1}
	{(\omega_{\lambda}+\omega_{\lambda'}+\omega_{\lambda''})_p}
	+
	\frac{ 6n_{\lambda} n_{\lambda''} - 3 n_{\lambda} n_{\lambda'} + 3n_{\lambda''}}
	{(\omega_{\lambda}+\omega_{\lambda'}-\omega_{\lambda''})_p}
	\right)
	%
	+9\Phi_{\lambda\bar{\lambda}\lambda''}\Phi_{\lambda'\bar{\lambda}'\bar{\lambda}''}
	\frac{4 n_{\lambda}( n_{\lambda'}+1)+1}
	{(\omega_{\lambda''})_p}\,,
\end{equation}
and
\begin{equation}\label{eq:deltaF4}
	\Delta F^{4\textrm{ph}} =
	3\sum_{\lambda\lambda'}
	\Phi_{\lambda\bar{\lambda}\lambda'\bar{\lambda}'}(2n_{\lambda}+1)(2n_{\lambda'}+1)
\end{equation}
The free energy was fitted to a Birch-Murnaghan equation of state, and pressure was determined via $P = -dF/dV$.

\subsection{Parameters in calculations}

All density functional theory calculations were done using the projector augmented wave (PAW)~\cite{Blochl1994a} method as implemented in VASP~\cite{Kresse1996c,Kresse1999,Kresse1996,Kresse1993b}. 
We treated exchange-correlation within the AM05 approximation~\cite{Armiento2005,Mattsson2009}. 
We used a 288 atom supercell, and the Brillouim zone integrations used a $2 \times 2 \times 2$ Monkhorst-Pack mesh,~\cite{Monkhorst1976a} and a Gaussian smearing of 0.2 eV was applied. 
Phonon self-energies and Raman spectra were integrated on a $51 \times 51 \times 51$ q-point mesh. 
The plane-wave energy cutoff was set to 350~eV.

The statistics needed to determine the polarizability and Born charges on a large set of supercells used a slightly smaller simulation cell to offset the increased computational cost. 
Here we used a 64 atom supercell with a tighter $6 \times 6 \times 6$ k-point mesh.

\newpage
\bibliography{main.bib}